\documentclass[a4paper, 11pt]{article}

\usepackage[utf8x]{inputenc}
\usepackage{epsfig,amssymb,euscript,xspace, color}
\usepackage{jheppub}
\usepackage[T1]{fontenc}
\usepackage{color}
\usepackage{amssymb}
\usepackage{amsfonts,amsmath,bm}
\usepackage{graphicx}
\usepackage{mathtools}
\usepackage{ragged2e}

\usepackage{dcolumn}% Align table columns on decimal point
\usepackage{graphicx}
\usepackage{braket}
\usepackage{verbatim}
\usepackage[english]{babel}
\usepackage{hyperref}
\hypersetup{
citecolor=red,
colorlinks=true,
filecolor=red,
linkcolor=blue,
linktocpage=true,
urlcolor=blue
}
\usepackage[titletoc,toc,title]{appendix}
\numberwithin{equation}{section}
\usepackage{cleveref}
\usepackage{epsfig}
\usepackage{float}
\usepackage{amsfonts}
\usepackage{enumitem}
\usepackage[font={footnotesize,it}]{caption}
\usepackage{cases}

\newcommand{\be}{\begin{equation}} 
\newcommand{\ee}{\end{equation}}
\newcommand{\eq}[1]{(\ref{#1})}
\newcommand{\bit}{\begin{itemize}}  \newcommand{\eit}{\end{itemize}}
\newcommand{\ben}{\begin{enumerate}}  \newcommand{\een}{\end{enumerate}}

\newcommand{\rf}[1]{(\ref{#1})}

\def\bd{\begin{document}}
\def\ed{\end{document}}
\def\bea{\begin{eqnarray}}
\def\eea{\end{eqnarray}}
\let\bm=\bibitem

\def\la{\langle}
\def\ra{\rangle}

\def\npb#1#2#3{Nucl. Phys. {\bf{B#1}} #3 (#2)}
\def\plb#1#2#3{Phys. Lett. {\bf{#1B}} #3 (#2)}
\def\prl#1#2#3{Phys. Rev. Lett. {\bf{#1}} #3 (#2)}
\def\prd#1#2#3{Phys. Rev. {D bf{#1}} #3 (#2)}
\def\cmp#1#2#3{Comm. Math. Phys. {\bf{#1}} #3 (#2)}
\def\cqg#1#2#3{Class. Quantum Grav. {\bf{#1}} #3 (#2)}
\def\nppsa#1#2#3{Nucl. Phys. B (Proc. Suppl.) {\bf{#1A}}#3 (#2)}
\def\ap#1#2#3{Ann. of Phys. {\bf{#1}} #3 (#2)}
\def\ijmp#1#2#3{Int. J. Mod. Phys. {\bf{A#1}} #3 (#2)}
\def\rmp#1#2#3{Rev. Mod. Phys. {\bf{#1}} #3 (#2)}
\def\mpla#1#2#3{Mod. Phys. Lett. {\bf A#1} #3 (#2)}
\def\jhep#1#2#3{J. High Energy Phys. {\bf #1} #3 (#2)}
\def\atmp#1#2#3{Adv. Theor. Math. Phys. {\bf #1} #3 (#2)}

\def\sst{\scriptscriptstyle}
\def\thetabar{\bar\theta}
\def\Tr{{\rm Tr}}
\def\one{\mbox{1 \kern-.59em {\rm l}}}

\def\a{\alpha}      \def\da{{\dot\alpha}}  \def\dA{{\dot A}}
\def\b{\beta}       \def\db{{\dot\beta}}
\def\g{\gamma}  \def\G{\Gamma}  \def\dc{{\dot\gamma}}
\def\d{\delta}  \def\D{\Delta}  \def\ddt{\dot\delta}
\def\e{\epsilon}
\def\ve{\varepsilon}
\def\uve{\upvarepsilon}
\def\f{\phi}    \def\F{\Phi}    \def\vvf{\f}
\def\vphi{\varphi}
\def\h{\eta}
\def\k{\kappa}
\def\l{\lambda} \def\L{\Lambda}
\def\m{\mu} \def\n{\nu}
\def\o{\omega}
\def\p{\pi} \def\P{\Pi}
\def\r{\rho}
\def\s{\sigma}  \def\S{\Sigma}
\def\t{\tau}
\def\th{\theta} \def\Th{\Theta} \def\vth{\vartheta}
\def\X{\Xeta}
\def\z{\zeta}

\def\na{\nabla}
\def\cA{{\cal A}} \def\cB{{\cal B}} \def\cC{{\cal C}}
\def\cD{{\cal D}} \def\cE{{\cal E}} \def\cF{{\cal F}}
\def\cG{{\cal G}} \def\cH{{\cal H}} \def\cI{{\cal I}}
\def\cJ{{\mathscr J}} \def\cK{{\cal K}} \def\cL{{\cal L}}
\def\cM{{\cal M}} \def\cN{{\cal N}} \def\cO{{\cal O}}
\def\cP{{\cal P}} \def\cQ{{\cal Q}} \def\cR{{\cal R}}
\def\cS{{\cal S}} \def\cT{{\cal T}} \def\cU{{\cal U}}
\def\cV{{\cal V}} \def\cW{{\cal W}} \def\cX{{\cal X}}
\def\cY{{\cal Y}} \def\cZ{{\cal Z}}
\def\ct{{\cal t}}

\def\ua{\underline{\alpha}}
\def\uc{\underline{\phantom{\alpha}}\!\!\!\gamma}
\def\um{\underline{\mu}}
\def\ud{\underline\delta}
\def\ue{\underline\epsilon}
\def\una{\underline a}\def\unA{\underline A}
\def\unb{\underline b}\def\unB{\underline B}
\def\unc{\underline c}\def\unC{\underline C}
\def\und{\underline d}\def\unD{\underline D}
\def\une{\underline e}\def\unE{\underline E}
\def\unf{\underline{\phantom{e}}\!\!\!\! f}\def\unF{\underline F}
\def\unm{\underline m}\def\unM{{\underline M}}
\def\unn{\underline n}\def\unN{{\underline N}}
\def\unp{\underline{\phantom{a}}\!\!\! p}\def\unP{\underline P}
\def\unq{\underline{\phantom{a}}\!\!\! q}
\def\unQ{\underline{\phantom{A}}\!\!\!\! Q}
\def\unH{\underline{H}}

\def\As {{A \hspace{-6.4pt} \slash}\;}
\def\bs {{b \hspace{-6.4pt} \slash}\;}
\def\Ds {{D \hspace{-6.4pt} \slash}\;}
\def\Gts {{\Gt \hspace{-6.4pt} \slash}\;}
\def\ds {{\del \hspace{-6.4pt} \slash}\;}
\def\ss {{\s \hspace{-6.4pt} \slash}\;}
\def\ks {{ k \hspace{-6.4pt} \slash}\;}
\def\ps {{p \hspace{-6.4pt} \slash}\;}
\def\xs {{x \hspace{-6.4pt} \slash}\;}
\def\pas {{{p_1} \hspace{-6.4pt} \slash}\;}
\def\pbs {{{p_2} \hspace{-6.4pt} \slash}\;}
\def\cFs {{{\cal F} \hspace{-6.4pt} \slash}\;}
\def\Dss {{D \hspace{-7.5pt} \slash}\;}
\def\dss {{\del \hspace{-7.0pt} \slash}\;}

\def\Ah{{\hat{A}}}
\def\Dh{{\hat{D}}}
\def\Gh{{\hat{G}}}
\def\Fh{{\hat{F}}}
\def\Ih{{\hat{I}}}
\def\Jh{{\hat{J}}}
\def\Kh{{\hat{K}}}
\def\Lh{{\hat{L}}}
\def\Ph{{\hat{P}}}
\def\Rh{{\hat{R}}}
\def\Vh{{\hat{V}}}
\def\Xh{{\hat{X}}}

\def\ah{{\hat{\a}}}
\def\bh{{\hat{\b}}}
\def\gh{{\hat{\g}}}
\def\dh{{\hat{\d}}}
\def\rh{{\hat{\r}}}
\def\hh{\hat{h}}
\def\uh{\hat{u}}
\def\xh{\hat{x}}
\def\yh{\hat{y}}
\def\ph{\hat{p}}
\def\xih{\hat{\xi}}
\def\chih{\hat{\chi}}
\def\Psih{\hat{\Psi}}
\def\phih{\hat{\phi}}

\def\psit{\tilde{\psi}}
\def\Psit{\tilde{\Psi}}
\def\Psibt{\tilde{\bar{Psi}}}

\def\lambdat{\tilde {\lambda}}
\def\st{\tilde{\sigma}}

\def\delt{\tilde{\delta}}
\def\Phit{\tilde{\Phi}}
\def\Phitb{\overline{\tilde{Phi}}}
\def\tht{\tilde{\th}}
\def\lt{\tilde{\l}}
\def\chit{\tilde{\chi}}
\def\phit{\tilde{\phi}}

\def\At{\tilde{A}}
\def\Bt{\tilde{B}}
\def\Ct{\tilde{C}}
\def\Dt{\tilde{D}}
\def\Et{\tilde{E}}
\def\Ft{\tilde{F}}
\def\Gt{\tilde{G}}
\def\Ht{\tilde{H}}
\def\It{\tilde{I}}
\def\Jt{\tilde{J}}
\def\Pt{\tilde{P}}
\def\Ot{\tilde{O}}
\def\Mt{\tilde{M }}
\def\Nt{\tilde{N}}
\def\St{\tilde{S}}
\def\Vt{\tilde{V}}
\def\Xt{\tilde{X}}
\def\at{\tilde{a}}
\def\dt{\tilde{d}}
\def\htt{\tilde{h}}
\def\ft{\tilde{f}}
\def\gt{\tilde{g}}
\def\pt{\tilde{p}}
\def\qt{\tilde{q}}
\def\rt{\tilde{r}}
\def\nt{\tilde{n}}
\def\ut{\tilde{u}}
\def\wt{\tilde{w}}
\def\zt{\tilde{z}}
\def\xt{\tilde{x}}
\def\yt{\tilde{y}}
\def\Psit{\tilde{\Psi}}
\def\phit{\tilde{\phi}}
\def\tD{\tilde{\D}}

\def\eb{\bar{\epsilon}}
\def\delb{\bar{\partial}}
\def\thb{\bar{\theta}}
\def\mub{\bar{\mu}}
\def\lamb{\bar{\l}}
\def\psib{\bar{\psi}}
\def\sb{\bar{\sigma}}
\def\xib{\bar{\xi}}
\def\chib{\bar{\chi}}

\def\Psib{\bar{\Psi}}
\def\Phib{\bar{\Phi}}
\def\Lamb{\bar{\Lambda}}
\def\Sb{{\overline \Sigma}}
\def\cb{\bar{c}}
\def\hb{\bar{h}}
\def\qb{\bar{q}}
\def\wb{\bar{w}}
\def\ub{\bar{u}}
\def\zb{{\bar{z}}}
\def\Hb{\bar{H}}
\def\Qb{{\bar Q}}
\def\Omegab{\overline{\Omega}}
\def\ob{\overline{\omega}}

\def\Ab{{\overline A}} \def\Bb{{\overline B}} \def\Cb{{\overline C}}
\def\Db{{\overline D}} \def\Eb{{\overline E}} \def\Fb{{\overline F}}
\def\Gb{{\overline G}}
\def\Ib{{\overline I}}
\def\Jb{{\overline J}} \def\Kb{{\overline K}} \def\Lb{{\overline L}}
\def\Mb{{\overline M}} \def\Nb{{\overline N}} \def\Ob{{\overline O}}
\def\Pb{{\overline P}}  \def\Rb{{\overline R}}
 \def\Tb{{\overline T}} \def\Ub{{\overline U}}
\def\Vb{{\overline V}} \def\Wb{{\overline W}} \def\Xb{{\overline X}}
\def\Yb{{\overline Y}} \def\Zb{{\overline Z}}

\def\fb{{\overline f}}
\def\gb{{\overline g}}
\def\nb{{\overline n}}
\def\mb{{\overline m}}
\def\lb{{\overline l}}
\def\yb{{\overline y}}

\def\ldel{{\overleftarrow{\del}}}
\def\rdel{{\overrightarrow{\del}}}
\def\ldeldel{{\overleftarrow{\del^2}}}
\def\rdeldel{{\overrightarrow{\del^2}}}
\def\ldelb{{\overleftarrow{\bar{\del}}}}
\def\rdelb{{\overrightarrow{\bar{\del}}}}
\def\ba{{\bf a}}
\def\bk{{\bf k}}
\def\bl{{\bf l}}
\def\bp{{\bf p}}
\def\bq{{\bf q}}
\def\br{{\bf r}}
\def\bt{{\bf t}}
\def\bu{{\bf u}}
\def\bv{{\bf v}}
\def\bx{{\bf x}}
\def\by{{\bf y}}
\def\bA{{\bf A}}
\def\bR{{\bf R}}
\def\bV{{\bf V}}

\def\bz{{\boldsymbol{\zeta}}}

\def\bone{{\bf 1}}

\def\va{{\vec a}}
\def\vk{{\vec k}}
\def\vp{{\vec p}}
\def\vq{{\vec q}}
\def\vx{{\vec x}}
\def\vy{{\vec y}}
\def\vu{{\vec u}}
\def\vv{{\vec v}}
\def \vH{{\vec H}}
\def \vg{{\vec g}}

\def\vs{{\vec \sigma}}
\def\vtau{{\vec \tau}}

\newcommand{\ov}[1]{\overrightarrow{#1}}

\def\frA{\mathfrak{A}}
\def\frB{\mathfrak{B}}
\def\frC{\mathfrak{C}}
\def\frD{\mathfrak{D}}
\def\frE{\mathfrak{E}}
\def\frF{\mathfrak{F}}
\def\frG{\mathfrak{G}}
\def\frH{\mathfrak{H}}
\def\frM{\mathfrak{M}}
\def\frN{\mathfrak{N}}
\def\frR{\mathfrak{R}}
\def\frW{\mathfrak{W}}

\def\fra{\mathfrak{a}}
\def\frb{\mathfrak{b}}
\def\frf{\mathfrak{f}}
\def\frg{\mathfrak{g}}
\def\frh{\mathfrak{h}}
\def\frl{\mathfrak{l}}
\def\frs{\mathfrak{s}}
\def\fri{\mathfrak{i}}
\def\frj{\mathfrak{j}}

\def\ma{\mathfrak{a}}
\def\mg{\mathfrak{g}}
\def\mh{\mathfrak{h}}
\def\mR{\mathfrak{R}}
\def\mN{\mathfrak{N}}

\newcommand{\nn}{{\nonumber}}

\def\d{\delta}\def\D{\Delta}\def\ddt{\dot\delta}

\def\pa{\partial} \def\del{\partial}
\def\xx{\times}
\def\uno{\mbox{1 \kern-.59em {\rm l}}}

\def\trp{^{\top}}
\def\inv{^{-1}}
\def\dag{\dagger}
\def\pr{^{\prime}}

\def\rar{\rightarrow}
\def\lar{\leftarrow}
\def\lrar{\leftrightarrow}

\newcommand{\0}{\,\!}      %this is just NOTHING!
\def\one{1\!\!1\,\,}
\def\im{\imath}
\def\jm{\jmath}

\newcommand{\tr}{\mbox{tr}}
\newcommand{\slsh}[1]{/ \!\!\!\! #1}

\newcommand{\1}{\mbox{1}\hspace{-0.25em}\mbox{l}}

\def\vac{|0\rangle}
\def\lvac{\langle 0|}

\def\hlf{\frac{1}{2}}
\def\ove#1{\frac{1}{#1}}
\newcommand{\hot}[1]{\frac{#1}{2}}

\def\Box{\square}
\def\CC {\mathbb{C}}
\def\FF {\mathbb{F}}
\def\RR{\mathbb{R}}
\def\NN{\mathbb{N}}
\def\ZZ{\mathbb{Z}}
\def\bb#1{{\bf #1}}
\def\bcomment#1{}
\def\bfhat#1{{\bf \hat{#1}}}
\def\VEV#1{\left\langle #1\right\rangle}

\newcommand{\ex}[1]{{\rm e}^{#1}} \def\ii{{\rm i}}

\newcommand{\lrbrk}[1]{\left(#1\right)}
\newcommand{\lrsbrk}[1]{\left[#1\right]}
\newcommand{\sfrac}[2]{{\textstyle\frac{#1}{#2}}}

\def\stw{{\sqrt{2}}}

\def\rf {{\rm f}}
\def\ri {{\rm i}}
\def\rj {{\rm j}}
\def\rn {{\rm n}}
\def\rk {{\rm k}}
\def\rl {{\rm l}}
\def\rr {{\rm r}}

\def\rQ {{\scriptscriptstyle \rm \cQ}}
\def\rR {{\scriptscriptstyle \rm \cR}}

\def\cQb{{\cal \Qb}}
\def\cRb{{\cal \Rb}}
\def\cWb{{\cal \Wb}}

\def\fd {{\rm N}}
\def\afd {{\overline{\rm N}}}

\def \II {I\hspace{-.1em}I\hspace{.1em}}
\def \IIA {\mbox{\II A\hspace{.2em}}}
\def \IIB {\mbox{\II B\hspace{.2em}}}
\def \gs {g^s}
\def \ls {\lambda^s}

\def \I {{\cal I}}
\def \qs {q\hspace{-.53em}/\hspace{.15em}}
\def \ks {k\hspace{-.53em}/\hspace{.15em}}
\def \YM {{\mbox{\tiny YM}}}
\def \gym {g_{\YM}}

\def \Lc {\L_c}
\def\IR{\relax{\rm I\kern-.18em R}}
\def \id {{\bf 1}}

\def\cci{\ell}
\def\ccj{\ell'}

\def\bbq{\pmb{q}}
\def\bom{\pmb{\o}}
\def\bJ{\pmb{J}}
\def\bM{\pmb{M}}
\def\bB{\pmb{B}}
\def\bn{\pmb{n}}
\def\bE{\pmb{E}}

\title{\Large \bf\boldmath Timelike entanglement entropy with gravitational anomalies}

\author[a,b,c]{Chong-Sun Chu}
\author[b,c]{Himanshu Parihar}

 \affiliation[a]{
	Department of Physics, National Tsing-Hua University,
 Hsinchu 30013, Taiwan}
\affiliation[b]{Center of Theory and Computation,
National Tsing-Hua University,
 Hsinchu 30013, Taiwan}
\affiliation[c]{Physics Division,
    National Center for Theoretical Sciences,
     Taipei 10617, Taiwan}

\emailAdd{cschu@phys.nthu.edu.tw}
\emailAdd{himansp@phys.ncts.ntu.edu.tw}

\abstract{\noindent We study the timelike entanglement entropy (TEE) in two dimensional conformal field theories (CFT) with gravitational anomalies. We employ analytical continuation to compute the timelike entanglement entropy for a pure timelike interval in such CFTs. We find that, unlike the real part, the imaginary part of the TEE displays an asymmetric dependence on the central charges of the left and right moving modes.
We propose that the asymmetric dependence on central charges of the imaginary part of the TEE can be used to probe the presence of gravitational anomalies in chiral CFT. Furthermore, we propose a holographic construction to obtain the timelike entanglement entropy from the bulk dual geometries involving topologically massive gravity in AdS$_3$. The holographic results obtained match exactly with the dual field theory results. }

\begin{document}
	
	\maketitle
	\flushbottom
	\pagebreak

	\definecolor{orange}{rgb}{1.0, 0.49, 0.0}

\section{Introduction}

The study of quantum entanglement in extended many body systems has
recently become a crucial area of research yielding insights across
condensed matter physics, quantum gravity and black hole physics. A
key tool in this investigation is entanglement entropy which
quantifies the entanglement in a given bipartite pure states.  For two
dimensional conformal field theories (CFTs), the replica technique
provides a means to calculate this entanglement entropy
\cite{Calabrese:2004eu,Calabrese:2009qy}. However when left and right
moving central charges in a CFT become unequal indicating a
gravitational anomaly \cite{Alvarez-Gaume:1983ihn}, the CFT exhibits a
breakdown of stress-energy conservation at the quantum level. This
anomaly arising from the non conservation of the stress-energy tensor
implies a coordinate dependent description of the theory. The authors
\cite{Castro:2014tta} obtained the entanglement entropy in two
dimensional conformal field theories with gravitational anomalies
using the replica technique and showed that it has an additional
contribution coming from the Lorentz anomaly. Alternative methods such
as the generalized Rindler approach have also been employed to compute
entanglement entropy in such CFTs \cite{He:2023cju}.  Subsequent
investigations have focused on analyzing other bipartite mixed state
entanglement measures such as reflected entropy, entanglement
negativity and balanced partial entanglement entropy (BPE) within
these anomalous CFTs \cite{Basu:2022nds,Wen:2022jxr}.

In the context of AdS/CFT, \cite{Castro:2014tta} provides a bulk
description for the holographic entanglement entropy in the presence
of a gravitational anomaly where the bulk dual is described by the
topologically massive gravity (TMG) in an asymptotically AdS$_3$
spacetime. The action for the TMG-AdS$_3$ consists of usual
Einstein-Hilbert term and a gravitational Chern-Simons term which
alters the motion of massive spinning particles propagating in the
bulk geometry. Their worldlines take the form of ribbons incorporating
a normal frame at each point. Due to its explicit dependence on the
connection, the Chern-Simons term is not manifestly invariant under
diffeomorphism. This absence of invariance directly results in unequal
central charges in the dual field theory. The contribution of the
Chern-Simons term to entanglement entropy is determined by the boost
needed to propagate the normal frame along the
worldline. Consequently, the holographic entanglement entropy is given
by the sum of the geodesic length (calculated using the RT formula
\cite{Ryu:2006bv,Ryu:2006ef}) and the twist of the normal frame along
the geodesic \cite{Castro:2014tta}. The holographic entanglement
entropy can also be determined using the Rindler method as discussed
in \cite{Jiang:2019qvd}. Recently, the correction to the holographic
entanglement entropy (also BPE) has also been reproduced through the
application of newly introduced surface termed inner RT (IRT) surface
\cite{Wen:2024muv}. See
\cite{Kraus:2005zm,Song:2016gtd,Jiang:2017ecm,Hijano:2017eii,Wen:2018mev,
  Gao:2019vcc,Song:2019txa,Sun:2008uf,Basu:2021axf,Wang:2023jfr}
for more development related to TMG-AdS$_3$/CFT$_2$ duality.

Previous research on entanglement entropy has primarily focused on
cases where the subsystem in question is a spacelike separated
region. However, recent studies have extended this to include timelike
scenarios. In this regard, the authors \cite{Doi:2022iyj,Doi:2023zaf}
introduced a new entanglement measure for timelike subsystems termed
\textit{timelike entanglement entropy}. It can be defined as the
analytic continuation of entanglement entropy of a spacelike subsystem
to a timelike one and exhibits complex values in a
CFT$_2$. Holographically as shown in \cite{Doi:2022iyj,Doi:2023zaf},
the timelike entanglement entropy (TEE) can be obtained by considering
combined spacelike and timelike extremal surfaces (geodesics) in the
bulk. Specifically, the spacelike extremal surface contributes the
real part of TEE while the timelike extremal surface contributes the
imaginary part which is obtained by embedding the Poincaré patch in
the global patch. Moreover it has been argued that TEE is a specific
example of pseudo entropy
\cite{Nakata:2020luh,Mollabashi:2020yie,Mollabashi:2021xsd}. 
The pseudo entropy is defined by replacing the reduced density matrix with a non-Hermitian transition matrix which generally results in a complex-valued entropy. Furthermore, the holographic pseudo entropy in dS/CFT has been shown to be related to the holographic TEE in AdS/CFT via a double Wick rotation as discussed in \cite{Doi:2022iyj}.
Since there can be many multiple combined spacelike and timelike surfaces
homologous to the boundary subsystem which gives infinitely many
different complex valued areas, \cite{Li:2022tsv} introduced the
complex-valued weak extremal surface (CWES) criteria which is a
complex valued generalization of the Ryu-Takayanagi extremal
surface. The unique surface is then selected by choosing the CWES with
the lowest imaginary area.  On the other hand, for geometries like
Lifshitz spacetime where a global embedding is not known, the
holographic computation for TEE must be performed entirely within the
Poincaré patch. As demonstrated in \cite{Basak:2023otu} by
extremization of length functional in Lifshitz metric, the holographic
TEE involves two spacelike geodesics connecting the timelike interval
endpoints to infinity and a timelike geodesic linking the spacelike
geodesic endpoints. The timelike geodesic is smoothly joined with the
spacelike geodesics at infinity. This method also reproduces the TEE
results in AdS$_3$/CFT$_2$ case \cite{Basak:2023otu}. Subsequently
this approach was extended to holographic TEE in non-conformal
theories and non-relativistic theories \cite{Afrasiar:2024lsi,
  Afrasiar:2024ldn}. In \cite{Heller:2024whi}, the authors obtain the
holographic TEE by suitably generalizing the Ryu-Takayanagi formula to
a complexified bulk geometry (refer to
\cite{Wang:2018jva,Liu:2022ugc,Guo:2022jzs,Narayan:2022afv,He:2023eap,Jiang:2023ffu,Narayan:2023ebn,Chu:2023zah,Jiang:2023loq,He:2023wko,Chen:2023eic,He:2023ubi,Narayan:2023zen,Shinmyo:2023eci,Guo:2023tjv,Diaz:2023npx,Kanda:2023jyi,He:2023syy,Das:2023yyl,Grieninger:2023knz,Guo:2024lrr,Basu:2024bal,Guo:2024edr,Anegawa:2024kdj,Goswami:2024vfl,Jena:2024tly,Xu:2024yvf,Roychowdhury:2025ukl,Guo:2025pru,Roychowdhury:2025aye}
for recent progress). Note that while the quantum mechanical interpretation of TEE is not yet understood properly but the AdS/CFT correspondence suggests a clearer perspective. Here the imaginary part of TEE is related to the area of a timelike extremal surface in the bulk. This implies that the imaginary part of TEE may be linked to the emergence of the time coordinate in holography, much like the entanglement entropy is associated with the emergence of spatial dimensions.
Very recently, a novel generalization of the
density matrix to capture correlations between timelike separated
subsystems have been introduced \cite{Milekhin:2025ycm} and
entanglement measures for timelike separated subsystems known as
entanglement in time was proposed. They further demonstrated that
entanglement in time coincides with the TEE for relativistic quantum
field theories. This equivalence also applies to the anomalous CFT
which we study here.

In this paper, we examine how gravitational anomalies affect TEE in
two dimensional CFTs considering both field theory and holographic
descriptions. In this context, we first compute TEE at zero
temperature, finite temperature with angular potential and finite
angular potential with zero temperature in a CFT$_2$ with a
gravitational anomaly.  This involves the analytic continuation of the
entanglement entropy of a spacelike interval to a timelike one. The
resulting TEE is complex and reveals distinct central charge
dependencies in its real and imaginary parts. Its imaginary part
depends on a single central charge while the real part on both left
and right central charges. We then compute TEE holographically,
showing that it involves summing on-shell actions of a massive
spinning particle moving on the spacelike and timelike extremal
worldlines. Given that the holographic TEE involves both spacelike and
timelike geodesics, and the Chern-Simons contribution for spacelike
geodesics is known from \cite{Castro:2014tta}, we here derive the
corresponding contribution for timelike geodesics. The key difference
is that the Chern-Simons contribution now quantifies the rotation of
normal frames along the timelike geodesic whereas it measures the
twist (or Lorentz boost) of normal vectors for spacelike
geodesics. This contribution from the Chern-Simons term exhibits a
complex nature similar to the length of a timelike geodesic. Since the
intersection of spacelike and timelike geodesics lie deep in the bulk,
we prescribe a specific boundary condition for normal vectors which is
based on the time direction in the boundary. Further these boundary
conditions are used to obtain the Chern-Simons (anomalous)
contribution from each type of geodesic to the TEE.  We show that
interestingly the bulk holographic results agree exactly with the dual
field theory results.

The rest of the paper is organized as follows. In \cref{sec-review} we
briefly review about the entanglement entropy in a CFT$_2$ with
gravitational anomaly. In \cref{sec-TEE} we compute the TEE of a pure
timelike interval using analytical continuation in a CFT$_2$ with a
gravitational anomaly. Subsequently in \cref{sec-HTEE} after reviewing
the holographic TMG-AdS$_3$/CFT$_2$ duality, we derive the anomaly
contribution due to timelike geodesic and propose a holographic
construction for the TEE. We then obtain the holographic TEE in a
CFT$_2$ dual to Poincar\'e AdS$_3$, non-extremal rotating BTZ black
hole and extremal rotating BTZ black hole. Finally in
\cref{sec-summary} we present our summary and discussions.

\section{Entanglement entropy in CFT$_2$ with gravitational anomalies}
\label{sec-review}

We first briefly review about the two dimensional CFT with a
gravitational anomaly
\cite{Alvarez-Gaume:1983ihn,Alvarez-Gaume:1984zlq}. Such theories are
described by two copies of the Virasoro algebra with two unequal
central charges $c_L$ and $c_R$, and there is a breakdown of
energy-momentum conservation at the quantum level.  The anomaly
appears as a gravitational anomaly where the stress tensor is
symmetric but not conserved. This non-conservation is captured by the
anomalous divergence of the stress tensor as \cite{Kraus:2005zm}
\begin{equation}
\nabla_\mu T^{\mu \nu}=\frac{c_L-c_R}{96 \pi} g^{\mu \nu} \epsilon^{\alpha \beta}
\partial_\alpha \partial_\rho \Gamma^\rho_{\nu \beta}.
\end{equation}
From a bulk perspective, gravitational anomalies arise because the
gravitational Chern-Simons term is only invariant under
diffeomorphisms up to a boundary term that leads to a non-zero
divergence of the dual field theory stress tensor.  Alternatively, the
anomaly can also arise from a broken Lorentz symmetry which is
anomalies under local frame rotations resulting in a conserved stress
tensor but not symmetric. This makes the theory frame dependent. One
can move between these two perspectives by incorporating a local
counterterm into the CFT generating functional
\cite{Alvarez-Gaume:1984zlq,Bardeen:1984pm}.

%-------------------------------------------------------

\subsection{Zero temperature}

\begin{figure}[H]
	\centering
	\includegraphics[scale=1.2]{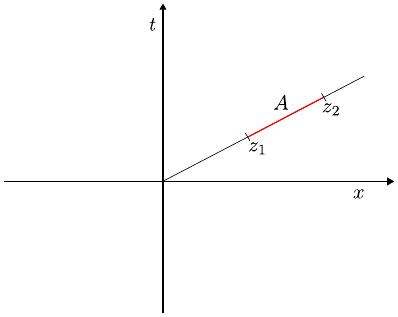}
	\caption{Configuration for a single boosted interval on a
          complex plane.}\label{single-interval}
\end{figure}

In this subsection we review the entanglement entropy computation for
a single interval at zero and finite temperature in a CFT$_2$ with a
gravitational anomaly as discussed in \cite{Castro:2014tta}.
Consider a boosted interval described by $A\equiv [z_1,z_2] =
[(x_1,t_1), (x_2,t_2)]$ and $B=A^c$ represent the
rest of the system. This configuration is depicted in \cref{single-interval}.

The entanglement entropy for a generic spacelike interval $A$ can be
expressed in terms of the two point twist field correlators as
\cite{Calabrese:2004eu,Calabrese:2009qy}
\begin{equation}
\begin{aligned}\label{S-def}
S_A=\lim_{n\to
  1}\frac{1}{1-n}\log\left<\Phi_{n}(z_1,\bar{z}_1)\Phi_{-n}(z_2,\bar{z}_2)\right>,
\end{aligned}
\end{equation}
where $\Phi_{n}(z_1,\bar{z}_1)$ and $\Phi_{-n}(z_2,\bar{z}_2)$ are the
twist and the anti twist fields with conformal dimensions
$h_L=\frac{c_L}{24}\left( n-\frac{1}{n}\right)$ and
$h_R=\frac{c_R}{24}\left( n- \frac{1}{n} \right)$. Since $c_L$ and
$c_R$ are unequal, the twist fields have a non-zero spin $s_n$
proportional to the difference in anomaly coefficients
$(c_L-c_R)$. The scaling dimension and spin of the twist fields are
then given by
\begin{align}\label{scaling-spin}
  \Delta_n=\frac{c_L+c_R}{24}\left(n-\frac{1}{n}\right)
  \,\,,\,\, s_n=\frac{c_L-c_R}{24}\left(n-\frac{1}{n}\right).
\end{align}
On using the two point twist correlator
\begin{equation}
    \left<\Phi_{n}(z_1,\bar{z}_1)\Phi_{-n}(z_2,\bar{z}_2)\right>=c_n
    z_{12}^{-2 h_L} \bar{z}_{12}^{-2 h_R},
\end{equation}
in \eqref{S-def}, one can obtain the entanglement entropy for
a generic spacelike interval $A$ as 
\begin{align}\label{EE-single}
  S_A=\frac{c_L}{6}\log \left(\frac{z_{12}}{\epsilon}\right)
  +\frac{c_R}{6}\log \left(\frac{\bar{z}_{12}}{\epsilon}\right),
\end{align}
where $\epsilon$ is a UV cut-off. Now analytically continuing to the
Lorentzian signature via $z=x-t$, $\bar{z}=x+t$ and using $z=R
e^{i\theta}$, the entanglement entropy can be rewritten as follows
\cite{Castro:2014tta}
\begin{equation}\label{spacelike-EE-zero}
  S_A=\frac{c_L+c_R}{6}\log\left( \frac{R}{\epsilon}\right)
  -\frac{c_L-c_R}{6}\kappa.
\end{equation}
where $\kappa$ is the boost parameter and related to rotation angle as
$\theta=i\kappa$. It is observed that the second term in the above
expression arises due to contribution from the gravitational anomaly.
The length $R$ and boost $\kappa$ for the boosted interval $A$ is
given in terms of $(t, x)$-coordinates as
\begin{align} \label{Length-boost}
R = \sqrt{x_{12}^2 - t_{12}^2} \, , \qquad \kappa = \tanh^{-1}\left(
\frac{t_{12}}{x_{12}} \right).
\end{align}
In the absence of an anomaly ($c_L=c_R$), the entanglement entropy
\eqref{spacelike-EE-zero} simplifies to the familiar entanglement
entropy of a single interval as described in
\cite{Calabrese:2004eu,Calabrese:2009qy}.

\subsection{Finite temperature and angular potential}

First note that the CFT$_2$ at finite temperature with the
gravitational anomaly is described by the grand canonical
ensemble. The unequal central charges give rise to a conserved spin
angular momentum whose conjugate chemical potential is known as the
angular potential.  Consider a generic spacelike interval $A\equiv
[w_1,w_2]$ in the CFT$_2$ at a finite temperature $T=\beta^{-1}$ and
with a non zero chemical potential $\Omega$ for the spin angular
momentum. The Euclidean partition function for this finite temperature
CFT$_2$ with a angular momentum $J$ as the conserved charge is given
by
\begin{equation}
Z=\text{Tr}\left(e^{-\beta\,H-\beta\Omega_EJ}\right),
\end{equation}
where $H$ is the Hamiltonian, $\beta$ is the inverse temperature, $J$
is the angular momentum and $\Omega_E = -i\Omega$ is the Euclidean
angular velocity. This state exhibits non vanishing energy and
momentum expectation values as follows
\begin{equation}
H=E_R+E_L-\frac{c_L+c_R}{24}\,, \quad J=E_R-E_L+\frac{c_L-c_R}{24}.
\end{equation}
The left and right moving inverse temperatures $(\beta_L,\beta_R)$
associated with the left and right moving modes in terms of
$(\beta,\Omega_E)$ is given by
\begin{equation}\label{tempsss}
\beta_L = \beta(1+i\Omega_E)~, \quad \beta_R = \beta(1-i\Omega_E)~.
\end{equation}
The CFT$_2$ for this case lives on a twisted cylinder due to the spin angular
momentum and is related to those defined on the complex plane through
the following conformal transformation
\begin{equation}\label{twisted-cylinder-transf}
z=e^{2\pi w/\beta_L}, \quad \bar{z}=e^{2\pi \bar{w}/\beta_R},
\end{equation}
where $(z,\bar{z})$ and $(w,\bar{w}$) are the coordinates on the
complex plane and the twisted cylinder respectively. The entanglement
entropy for an interval $A$ may now be obtained by computing the two
point twist correlator on the twisted cylinder using the above
conformal transformation as
\begin{equation}\label{finite-T-entropy-LR}
S_A= \frac{c_L}{6} \log \left [ \frac{\beta_L}{\pi \epsilon} \sinh
  \frac{\pi w_{12}}{\beta_L}\right] + \frac{c_R}{6} \log \left [
  \frac{\beta_R}{\pi \epsilon} \sinh \frac{\pi \bar{w}_{12}}{\beta_R}
  \right ],
\end{equation}
where $w_{12}=x_{12}+i\tau_{12}$ and $\bar{w}_{12}=x_{12}-i\tau_{12}$
are the interval lengths of $A$ on the twisted cylinder. The above
expression factorizes into the entanglement entropy for left and right
moving temperatures. To see a contribution from the anomaly, on
rearranging \eq{finite-T-entropy-LR} and analytically continuing to
Lorentzian signature via $\tau\to it$, the entanglement entropy for an
interval $A$ at finite temperature with a conserved angular momentum
is given by

\begin{equation}\label{entropy-single-finite}
S_A= \frac{c_L+c_R}{12} \log \left [ \frac{\beta_L \beta_R}{\pi^2
    \epsilon^2} \sinh \frac{\pi (x_{12}-t_{12})}{\beta_L} \sinh
  \frac{\pi (x_{12}+t_{12})}{\beta_R}\right ] - \frac{(c_L-c_R)}{12}
\log \left [\frac{\beta_R \sinh \frac{\pi (x_{12}+t_{12})}{\beta_R}
  }{\beta_L \sinh \frac{\pi (x_{12}-t_{12})}{\beta_L} }\right ].
\end{equation}
Here the second term represents the contribution from gravitational
anomaly. The above expression reduces to the entanglement entropy of
an interval in a CFT$_2$ at finite temperature with an angular
potential \cite{Hubeny:2007xt} in the absence of an anomaly.

\subsection{Finite angular potential and zero temperature}

Consider a generic spacelike interval $A\equiv [w_1,w_2]$ in a CFT$_2$
at a finite angular potential and zero temperature. For this case one
of the inverse temperatures i.e $\beta_L$ goes to infinity and the
other one $\beta_R$ remains finite. The conformal transformation
relating correlation functions on the complex plane to those on the
twisted cylinder is given by\footnote{Since CFT$_2$'s with a conserved
angular momentum at zero temperature are dual to the extremal rotating
BTZ black holes, the corresponding conformal transformation can be
obtained by taking the limit $r\to \infty$ in \eq{Coord-extremal}
\cite{Caputa:2013lfa}.}

\begin{equation}
    z=w, \quad \quad \bar{z}=\frac{\ell}{2 r_0}e^{2\,r_0\bar{w}/\ell}.
\end{equation}
By utilizing the above conformal transformation, the entanglement
entropy for an interval $A$ can be obtained by evaluating the
two-point twist correlator on the twisted cylinder as
\cite{Caputa:2013lfa}
\begin{equation}\label{entropy-single-finite-potential-zero-T}
S_A = \frac{c_L}{6} \log \frac{w_{12}}{\epsilon} + \frac{c_R}{6} \log
\left( \frac{\ell}{r_0 \epsilon} \sinh \frac{r_0
  \bar{w}_{12}}{\ell}\right),
\end{equation}
where $w_{12}=x_{12}+i\tau_{12}$ and
$\bar{w}_{12}=x_{12}-i\tau_{12}$. As described in
\cite{Caputa:2013lfa}, the first part of the above equation is the
entanglement entropy of an interval $A$ at zero temperature and the
second part is the entanglement entropy at an effective temperature
$r_0/\pi \ell$. This structure arises because the left moving modes of
the CFT are in their ground state while the right moving modes are at
an effective temperature known as the Frolov-Thorne temperature
$T_{FT}=r_0/\pi \ell$ \cite{PhysRevD.39.2125}. To see the anomaly
contribution explicitly, upon rearranging the above expression and
analytically continuing to the Lorentzian signature leads to the
entanglement entropy of an interval at zero temperature with finite
angular momentum as follows
\begin{equation}\label{EE-finite-potential-zero-T}
S_A= \frac{c_L+c_R}{12} \log \left [ \frac{\ell (x_{12}-t_{12})}{r_0
    \epsilon^2} \sinh \frac{r_0(x_{12}+t_{12})}{\ell}\right ] -
\frac{(c_L-c_R)}{12} \log \left [ \frac{\ell}{r_0
    (x_{12}-t_{12})}\sinh \frac{r_0(x_{12}+t_{12})}{\ell} \right ].
\end{equation}
The second term in the above expression represents contribution due to
the gravitational anomaly.

\section{Timelike entanglement entropy in CFT$_2$ with gravitational anomaly}
\label{sec-TEE}

We now proceed to compute the timelike entanglement entropy (TEE) of
an interval in a CFT$_2$ with unequal central charges or gravitational
anomaly involving different scenarios i.e at zero temperature, finite
temperature with angular potential and finite angular potential with
zero temperature. This involves the analytic continuation of the
entanglement entropy for a spacelike subsystem to a timelike
subsystem.

\subsection{Zero temperature}

The timelike entanglement entropy for a timelike interval $A$ in
CFT$_{2}$ in the presence of gravitational anomaly is obtained by
analytically continuing \eq{spacelike-EE-zero} to the timelike
interval case where $t_{12}^2-x_{12}^2>0$. On doing so,
\eq{spacelike-EE-zero} becomes
\begin{equation}\label{TEE-gen-zero-T}
S_A^T=\frac{c_L+c_R}{6}\log
\frac{\sqrt{t_{12}^2-x_{12}^2}}{\epsilon}+\frac{c_L+c_R}{12}
\log (-1) -\frac{c_L-c_R}{6}\tanh^{-1}\left(\frac{t_{12}}{x_{12}}\right).
\end{equation}
Here the function $\tanh^{-1}z$ is defined  in terms of the complex logarithm as
\begin{equation} \label{tanh}
    \tanh^{-1}z=\frac{1}{2}\log \frac{1+z}{1-z}.
\end{equation}
Now for a pure timelike interval, we set $x_{12}=0$ and $t_{12}=T$ in \eqref{TEE-gen-zero-T}
to obtain the timelike entanglement entropy as\footnote{Since for a pure
timelike interval $\frac{t_{12}}{x_{12}}\to \infty$
as $x_{12} \to 0^+$ and $t_{12}>0$,
one obtain from \eq{tanh}
\begin{equation}
  \tanh^{-1}\infty
    = \frac{1}{2}\log (-1).
\end{equation}
}
\begin{equation}
  S_A^T=\frac{c_L+c_R}{6}\log \frac{T}{\epsilon}+\frac{c_L+c_R}{12} \log(-1)
  -\frac{c_L-c_R}{12}\log (-1).
\end{equation}
We will consider the principal value of log from now on, then the timelike
entanglement entropy in the presence of anomaly for a pure timelike
interval is given by
\begin{subequations}\label{TEE-zero-temp}
\begin{align}
    S_A^T&=\frac{c_L+c_R}{6}\log \frac{T}{\epsilon}+\frac{c_L+c_R}{12}
    i \pi-\frac{c_L-c_R}{12}i
    \pi\label{TEE-a}\\ &=\frac{c_L+c_R}{6}\log
    \frac{T}{\epsilon}+\frac{c_R}{6}i \pi. \label{TEE-b}
\end{align}
\end{subequations}
This result  \eq{TEE-b} generalizes that of
\cite{Doi:2022iyj,Doi:2023zaf} for the timelike
entanglement entropy in non-anomalous CFT. In the presence of
gravitational anomaly, the TEE 
acquires an additional contribution proportional to the anomaly
coefficient $c_L-c_R$ (third term in \eq{TEE-a}) and
a partial cancellation of the usual imaginary part
of the TEE with the contribution from the gravitational anomaly occurs.
This results in an asymmetric
dependence of the TEE on the central charges,
with the real part depends on both central charges
and the imaginary part depends only on one of the two central charges.
In particular,
the TEE of the left moving mode is purely
real. This later property
is an useful feature that one may exploit to detect the presence of
gravitational anomaly in a chiral CFT.

\subsection{Finite temperature and angular potential}

We now compute the timelike entanglement entropy in a two dimensional
CFT at finite temperature and non-zero chemical potential. The
timelike entanglement entropy is obtained by analytically continuing
the spacelike interval $A$ in \eq{entropy-single-finite} to the
timelike interval by making $t^2_{12}>x^2_{12}$ as
\begin{equation}
\begin{aligned}
    S_A^T&=\frac{c_L+c_R}{12} \log \left [ \frac{\beta_L
        \beta_R}{\pi^2 \epsilon^2} \sinh \frac{\pi
        (t_{12}-x_{12})}{\beta_L} \sinh \frac{\pi
        (t_{12}+x_{12})}{\beta_R}\right ] - \frac{(c_L-c_R)}{12} \log
    \left [ \frac{\beta_R \sinh \frac{\pi (t_{12}+x_{12})}{\beta_R}
      }{\beta_L\sinh \frac{\pi (t_{12}-x_{12})}{\beta_L} }\right
    ]\\ &+\frac{c_L+c_R}{12}i \pi 
    -\frac{(c_L-c_R)}{12}i \pi.
\end{aligned}
\end{equation}
Then the timelike entanglement entropy for a pure timelike interval at
finite temperature and angular potential can be obtained by taking
$x_{12}=0$ and $t_{12}=T$ in the above expression as follows
\begin{equation}\label{TEE-finite-T-A}
S_A^T=\frac{c_L+c_R}{12} \log \left [ \frac{\beta_L \beta_R}{\pi^2
    \epsilon^2} \sinh \frac{\pi T}{\beta_L} \sinh \frac{\pi
    T}{\beta_R}\right ] - \frac{(c_L-c_R)}{12} \log \left [
  \frac{\beta_R\sinh \frac{\pi T}{\beta_R} }{\beta_L\sinh \frac{\pi
      T}{\beta_L} }\right ]+\frac{c_R}{6}i \pi.
\end{equation}
We find that the above TEE is complex where only one central charge
contributes to the imaginary part and both central charges are present
in the real part, similar to the zero temperature case. However the
anomaly contribution
to the
real part (proportional to $c_L-c_R$) is
present in contrast to earlier case. It matches with the CFT$_2$
result obtained in \cite{Doi:2023zaf} when the anomaly is absent
i.e. $c_L=c_R$.
We note that in the zero temperature limit
$(\beta_L,\beta_R)\gg T$, the above expression reduces to
\eq{TEE-zero-temp}.
We also note that one can consider the Rindler
approach  \cite{He:2023ubi} in the presence of gravitational anomaly
and obtain the same results \eq{TEE-finite-T-A} and \eq{TEE-zero-temp}.

\subsection{Zero temperature and finite angular potential}

Next, we obtain the timelike entanglement entropy of a pure timelike
interval in a CFT$_2$ at zero temperature characterized by finite
angular momentum. This can be computed by the analytical continuation
of a spacelike interval $A$ to the timelike one followed by taking
$x_{12}=0$ and $t_{12}=T$ in \eq{EE-finite-potential-zero-T} as
follows
\begin{equation}\label{TEE-zero-T-finite-ap}
S_A^T= \frac{c_L+c_R}{12} \log \left [ \frac{\ell \,T}{r_0 \epsilon^2}
  \sinh \frac{r_0 T}{\ell}\right ] - \frac{(c_L-c_R)}{12} \log \left [
  \frac{\ell}{r_0 T}\sinh \frac{r_0 T}{\ell} \right
]+\frac{c_R}{6}i\pi.
\end{equation}
We observe that,
like all the previous cases,
the TEE for this case is complex with only one
central charge
appearing in the imaginary part and both central
charges appearing in the real part.

\section{Holographic Timelike entanglement entropy in AdS$_3$/CFT$_2$ with gravitational anomaly}\label{sec-HTEE}

In this section, we compute the timelike entanglement entropy
holographically in the context of the AdS/CFT correspondence for dual
conformal field theories with a gravitational anomaly. In this case
the bulk dual geometry is described by topologically massive gravity
(TMG) in a bulk AdS$_3$ spacetime \cite{Kraus:2005zm, Castro:2014tta}.

\subsection{Review of earlier works}\label{TMG-review}

We first briefly review about the holographic description in the
context of Topologically Massive Gravity (TMG) in AdS$_3$. The bulk
action for TMG in AdS$_3$ consists of the standard Einstein-Hilbert
action and an additional gravitational Chern-Simons (CS) term which is
given by \cite{Tachikawa:2006sz,Skenderis:2009nt,Castro:2014tta}
\begin{equation}\label{TMG-action}
S_{\rm TMG}={1\over 16\pi G_N^{(3)}}\int
d^3x\sqrt{-g}\left(R+{2\over\ell^2}\right) +\frac{1}{32\pi
  G_N^{(3)}\mu} \int
d^3x\sqrt{-g}\epsilon^{\lambda\mu\nu}\Gamma^\rho_{\lambda\sigma}\left(\partial_\mu
\Gamma^\sigma_{\rho\nu}+{2\over
  3}\Gamma^\sigma_{\mu\tau}\Gamma^\tau_{\nu\rho}\right), 
  \end{equation}
where $\mu$ is a parameter of mass dimension governing the coupling of the
CS term and $\ell$ represents the AdS$_3$ radius. The equations of
motion derived from this action take the following form
\begin{equation}\label{EOM-TMG} 
R_{\mu\nu}-{1\over 2}g_{\mu\nu} R -{1\over \ell^2}
g_{\mu\nu}=-{1\over \mu}C_{\mu\nu}, 
\end{equation}
where $C_{\mu\nu}$ denotes the
Cotton tensor. When the Cotton tensor vanishes, the theory admits
solutions that retain the form of Einstein metrics ensuring that these
solutions remain locally AdS$_3$. In this work, we confine our
discussion to such locally AdS$_3$ solutions.  The asymptotic symmetry
analysis of TMG in AdS$_3$ reveals that the algebra of asymptotic
Killing vector modes corresponds to two copies of the Virasoro algebra
with left and right moving central charges. This results in a
Brown-Henneaux \cite{Brown:1986nw} type relation for TMG-AdS$_3$ given
by \cite{Hotta:2008yq,Compere:2008cv}
\begin{equation}\label{BH-formula}
c_L={3\ell\over 2G_N^{(3)}}\left(1-{1\over \mu\ell}\right) ~,
\quad c_R= {3\ell\over 2G_N^{(3)}}\left(1+{1\over\mu\ell}\right).
\end{equation}
This relation suggests that the dual conformal field theory
exhibits a gravitational anomaly.

\subsubsection*{Holographic entanglement entropy in TMG-AdS$_3$}

For AdS/CFT correspondence, the authors \cite{Castro:2014tta} proposed
a bulk description for the holographic entanglement entropy in the
presence of gravitational anomalies by extending the
Lewkowycz-Maldacena prescription \cite{Lewkowycz:2013nqa}.  They
showed that the holographic entanglement entropy in such cases can be
obtained by extremizing a worldline action for a spinning particle
propagating in the AdS$_3$ bulk. Consequently, the holographic
entanglement entropy is given by the on-shell action of a massive
spinning particle in the topologically massive gravity as
\begin{equation}\label{EE-general}
S_{\rm on-shell} = \frac{1}{4G_N^{(3)}}\int_{C} ds
\left(\sqrt{g_{\mu\nu} \dot{X}^{\mu}\dot{X}^{\nu}} + \frac{i}{\mu} n_2
\cdot \nabla n_1 \right)_E,
\end{equation}
where $C$ represents the worldline of the particle. Here $n_1$ and
$n_2$ are the normal vectors to the curve (up to a choice of
handedness) such that $n_1\cdot n_2=0$ and $\nabla$ denotes the covariant derivative along the worldline which is defined as
\begin{equation}
\nabla V^\mu \equiv \frac{d V^\mu}{ds}+\Gamma^{\mu}_{\lambda\rho}\frac{d X^\rho}{ds} V^\lambda .
\end{equation}
The subscript $E$ indicates
that the above expression is evaluated in the Euclidean signature. For locally
AdS$_3$ spacetimes where the worldline that minimizes the action is a
geodesic, the first part in the above equation corresponds to the
Ryu-Takayanagi term originating from the Einstein-Hilbert action and
the second part arises from the Chern-Simons term.  

When the path (curve) $C$ is spacelike, one can choose
$n_1=\partial_t$ and $n_2=\partial_x$ where they are both spacelike in
the Euclidean signature. On analytically continuing to the complex
coordinates $(z,\bar{z})$ via $z = x - t, \bar{z} = x + t$, the normal
vectors $(n, \tilde{n})$ in Lorentzian signature is given by
\begin{equation}\label{spacelike-normal-vector-bdy}
n :=i n_1 = \partial_t \qquad \tilde{n} := n_2 = \partial_x .
\end{equation}
In Lorentzian signature, one of the normal vector becomes timelike
$(n)$ and the other remains spacelike $(\tilde n)$. The entanglement
entropy \eq{EE-general} in terms of these normal vectors can be
written as
\begin{equation}\label{spacelike-EE}
S_{\rm on-shell}^{\rm spacelike} = \frac{1}{4G_N^{(3)}}\int_{C} ds
\left(\sqrt{g_{\mu\nu} \dot{X}^{\mu}\dot{X}^{\nu}} + \frac{1}{\mu}
\tilde{n} \cdot \nabla n \right).
\end{equation}
The above expression gives the on-shell action for the case when the
path $C$ is spacelike. Here $s$ parametrizes the length along the path
(worldline) $C$ of the particle, $n$ and $\tilde{n}$ are unit timelike
and spacelike vectors respectively, both normal to the trajectory of
the particle $X ^\mu$. The second term in \eqref{spacelike-EE}
corresponds to the Chern-Simons contributions and denoted by $S_{\rm
  anom}^{\rm spacelike}$ can be written in terms of a single normal
vector $n$ using $\tilde{n}^{\mu} = \epsilon^{\mu\nu\rho} v_{\nu}
n_{\rho}$ as
\begin{equation}\label{space-like-anomaly-generic}
S_{\rm anom} = \frac{1}{4 G_N^{(3)} \mu} \int_C ds\;
\epsilon_{\mu\nu\rho}v^{\mu} n^{\nu}(\nabla n^{\rho}),
\end{equation}
where $v^\mu=\dot{X}^\mu$ is the tangent vector to the path $C$. The
above integral is insensitive to the variation of $n$ along the path
$C$, and depends only on the initial and the final value of the normal
vector $n$ ($\tilde{n}$). The boundary values of $n$ are denoted by
\be\label{n-BC}
n(s_i) = n_i ,\quad n(s_f) = n_f,
\ee
where $n_i$ and
$n_f$ are normal vectors at the end point of the path (or geodesic).
As described in \cite{Castro:2014tta}, the integral
\eq{space-like-anomaly-generic} measures the twist of $n_f$ relative
to $n_i$ along $C$. To obtain a compact expression for this integral,
one can introduce a vector $n(s)$ that obeys the boundary condition
\eq{n-BC} and a normal frame along the path defined by the pair of
vectors $(q^{\mu}, \tilde{q}^{\mu})$ which are both parallel
transported i.e. $\nabla q = \nabla \tilde{q} = 0$.  The normal vector
$n(s)$ can then be written in terms of $q$ and $\tilde{q}$ as
\be\label{n-eta}
n(s) = \cosh(\eta(s)) q(s) + \sinh(\eta(s))
\tilde{q}(s),
\ee
where $\eta$ is the rapidity of the Lorentz boost
required to boost the orthonormal frame $(v,q,\tilde{q})$ to
$(v,n,\tilde{n})$. The integral \eq{space-like-anomaly-generic} then
simplifies to a total derivative given by
\be\label{anom-difference}
S_{\rm anom}^{\rm spacelike} = \frac{1}{4G_N^{(3)} \mu} \int ds\,
\dot\eta(s) = \frac{1}{4 G_N^{(3)}\mu} \left(\eta(s_f) -
\eta(s_i)\right).
\ee
The anomalous contribution from the spacelike
curve can be written in terms of $q$ and $\tilde{q}$ by utilizing
\eqref{n-eta} and \eq{anom-difference} as \cite{Castro:2014tta}
\begin{equation}\label{spacelike-anomalous-formula}
S_{\rm anom}^{\rm spacelike} = \frac{1}{4G_N^{(3)}
  \mu}\log\left(\frac{q(s_f) \cdot n_f - \tilde{q}(s_f) \cdot
  n_f}{q(s_i) \cdot n_i - \tilde{q}(s_i) \cdot n_i}\right).
\end{equation}
Here $s$ parametrizes the spacelike geodesic and the vectors
$(q(s),\tilde{q}(s))$ describes a reference parallel transported
normal frame satisfying the following condition
\be\label{spacelike-condition}
\begin{aligned}
v^2= \tilde{q}^2=1& ,\quad q^2=-1 ,\quad q \cdot \tilde{q} =v\cdot q
=v\cdot \tilde{q} =0\\ &\nabla v = 0\ , \quad \nabla q=0\ , \quad
\nabla \tilde{q} = 0.
\end{aligned}
 \ee The holographic entanglement entropy is determined only by
 spacelike geodesic in the bulk and hence, it is obtained by
 extremizing the on-shell action \eq{spacelike-EE} as follows
 \cite{Castro:2014tta}
\begin{equation}
\begin{aligned}
    S_{\rm HEE}=\mathrm{min}_C\,\,\mathrm{ext} \, S_{\rm
      on-shell}^{\rm spacelike} =\mathrm{min}_C\,\,\mathrm{ext}
    \,\,(S_{\rm EH}^{\rm spacelike}+S_{\rm anom}^{\rm spacelike}),
    \end{aligned}
\end{equation}
which is the sum of length of the geodesic coming from the
Einstein-Hilbert term and the twist from the Chern-Simons
contribution.

\subsection{Anomalous contribution from timelike geodesic}

Having described the anomalous contribution from the spacelike
geodesic to the holographic entanglement entropy, we now proceed to
derive the anomalous (or CS) contribution from timelike geodesic to
the on-shell action \eq{EE-general}.  For a timelike path (or
geodesic) denoted as $C_{\tau}$, the tangent vector is timelike and
the other two normal vectors are spacelike which is different from the
spacelike geodesic case. We now choose the normal vectors $(n,
\tilde{n})$ in \eq{EE-general} as
\begin{equation}
n := n_1 =  \partial_u \qquad \tilde{n} := n_2 = \partial_x .
\end{equation}
These vectors remain spacelike under analytic continuation to
the Lorentzian signature. We now find the following expression for the
on-shell action for a timelike curve in terms of these vectors in the
Lorentzian signature using \eq{EE-general} as
\be\label{timelike-onshell-action}
S_{\rm on-shell}^{\rm timelike} = \frac{i}{4G_N^{(3)}}\int_{C_\tau}
d\tau \left(\sqrt{g_{\mu\nu} \dot{X}^{\mu}\dot{X}^{\nu}}
+ \frac{1}{\mu} \tilde{n} \cdot \nabla n \right),
\ee
where $\tau$ parameterizes the timelike curve $C_\tau$. We see that
\eq{timelike-onshell-action} becomes imaginary as length of timelike
curve is imaginary and also anomalous term becomes imaginary. Since
the first term in \eq{timelike-onshell-action} can be obtained by
computing the geodesic length, we now focus on the second term denoted
by $S^{\rm timelike}_{\rm anom}$.  It can be written in terms of a
single normal vector $n$ using $\tilde{n}^{\mu} =
\epsilon^{\mu\nu\rho} v_{\nu} n_{\rho}$ as
\begin{equation}\label{anom-single-vector}
  S_{\rm anom}^{\rm timelike} = \frac{i}{4 G_N^{(3)} \mu} \int_{C_{\tau}}
  d\tau\; \epsilon_{\mu\nu\rho}v^{\mu} n^{\nu}(\nabla n^{\rho}).
\end{equation}
Similar to the earlier section, the integration only depend on the
boundary terms i.e. it depends only on the initial and the final
values of the normal vector $n$ (or $\tilde{n}$). We set the boundary
values of the normal vector $n(\tau)$ as
\begin{equation}\label{bc-timelike}
n(\tau_i) = n_i ,\quad n(\tau_f) = n_f,
\end{equation}
where $n_i$ and $n_f$ are normal vectors located at the end point of
the curve.  As both the normal vectors ($n,\tilde{n}$) are spacelike
for this case, they rotate along the curve as we parallel transport
them from initial to the final point. In other words when we parallel
transport $n_i$ to the final point, the resulting vector is related to
$n_f$ via $SO(2)$ transformation.\footnote{They are related via
$SO(1,1)$ transformation for the case of spacelike curves where one
normal vector is timelike and other is spacelike
\cite{Castro:2014tta}.}  Now proceeding similar to spacelike curve
case, consider a vector $n(\tau)$ that satisfies the boundary
condition \eq{bc-timelike} and a parallel transported normal frame
which consists of two vectors $(q^{\mu}, \tilde{q}^{\mu})$ satisfying
$\nabla q = \nabla \tilde{q} = 0$. We can expand $n(\tau)$ in terms of
$q$ and $\tilde{q}$ as
\begin{equation}\label{n-q-tee}
n(\tau) = \cos(\theta(\tau)) q(\tau) - \sin(\theta(\tau)) \tilde{q}(\tau).
\end{equation}
Now using \eq{n-q-tee} in \eq{anom-single-vector} leads to
\begin{equation}\label{anom-theta}
  S_{\rm anom}^{\rm timelike} = \frac{i}{4G_N^{(3)} \mu} \int d\tau\,
  \dot\theta(\tau) = \frac{i}{4 G_N^{(3)}\mu} \left(\theta(\tau_f)
  - \theta(\tau_i)\right) .
\end{equation}
We observe that the anomaly contribution measures the rotation of
frames along the timelike curve which is different from spacelike case
where it measures the twist or Lorentz boost along the curve.  On
using \eqref{n-q-tee} in \eq{anom-theta}, the anomaly contribution can
be obtained in terms of $q$ and $\tilde{q}$ as
\begin{equation}\label{timelike-anomalous-formula}
S_{\rm anom}^{\rm timelike} = \frac{1}{4G_N^{(3)} \mu}\log \left[\frac{q(\tau_f)\cdot n_f-i \,\tilde{q}(\tau_f)\cdot n_f}{q(\tau_i)\cdot n_i-i \,\tilde{q}(\tau_i)\cdot n_i}\right].
\end{equation}
Although the angle of rotation of normal frames i.e $\int_{C_\tau}
\tilde{n} \cdot \nabla n$ along the timelike geodesic is real, the
overall anomaly contribution is complex for this case. The parallel
transported vectors used above satisfy the following condition
\begin{equation}\label{Timelike-q-condition}
 \begin{aligned}
   v^2= -1&,\quad q^2=\tilde{q}^2=1,\quad q \cdot
   \tilde{q} =v\cdot q =v\cdot \tilde{q} =0\\
&\nabla v = 0\ , \quad \nabla q=0\ , \quad \nabla \tilde{q} = 0.
 \end{aligned}
\end{equation}
As holographic TEE involves both spacelike and timelike geodesics, it
is determined by the total of the on-shell actions described by
\eq{spacelike-EE} and \eq{timelike-onshell-action} as follows
\begin{equation}\label{TEE-total-expression}
  S_{\rm TEE}=\mathrm{ext}_C ~
  \left(S_{\rm on-shell}^{\rm spacelike}+S_{\rm on-shell}^{\rm timelike}\right).
\end{equation}
This expression incorporates the geodesic length (RT formula) along
with the anomalous contributions arising from Chern-Simons term
considering both spacelike and timelike geodesic contributions. We now
utilize this to determine the holographic TEE in the following
section.

\subsection{Holographic TEE at zero temperature}

\begin{figure}[H]
	\centering
	\includegraphics[scale=1.2]{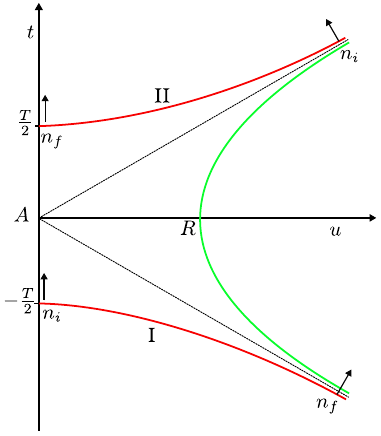}
	\caption{RT surfaces (geodesics) for a pure timelike interval
          in Poincar\'e coordinates are labelled by red and green
          curves. The normal vectors $n_i$ and $n_f$ are timelike for
          spacelike geodesics.}\label{spacelike-normal}
\end{figure}

Let us consider a pure timelike interval $A$ of length $T$ described
by $A\equiv[-\frac{T}{2},\frac{T}{2}]$ in a dual CFT$_2$ with a
gravitational anomaly at zero temperature. The corresponding
holographic dual is given by the AdS$_3$ spacetime in Poincar\'e
coordinates as
\be\label{ Poincar\'e-metric}
ds^2={\ell^2\over
  u^2}(-dt^2 + dx^2 +du^2),
\ee
where $\ell$ is the AdS radius.  As
discussed in \cite{Basak:2023otu} for the case of AdS$_3$/CFT$_2$, the
geodesics required for computing the holographic TEE consists of two
spacelike geodesics connecting the endpoints of $A$ and infinities
plus a timelike geodesic which connects the endpoints of two spacelike
geodesics. The equation for these geodesics in the Poincar\'{e} metric
\eq{ Poincar\'e-metric} is given by
	\begin{subequations}\label{geodesic-zero-T}
		\begin{align}
		   t^2-u^2&=\frac{T^2}{4}
                  \label{spacelike-geod}\\
		   u^2-t^2&=R^2,
                  \label{timelike-geod}
			\end{align}
		\end{subequations}
where $T$ and $R$ are constants. The curve \eq{spacelike-geod} is
spacelike and describes a geodesics that starts from the end point
$t=\pm T/2$ and ends at infinity. These spacelike geodesics are
smoothly joined at infinity by a timelike geodesic \eq{timelike-geod}
for any $R>0$. It is depicted in \cref{spacelike-normal}. 

In the presence of an anomaly, the holographic TEE acquires an
additional contribution due to the Chern-Simons term present in the
bulk gravitational action. Since the holographic TEE consists of both
spacelike and timelike geodesics, we must compute the contribution
from each geodesic separately. This involves computing the geodesic
length and the twist (or rotation) of the normal vectors along each
type of geodesic as described in the following subsections.

\subsection*{Spacelike geodesic contribution}

The equation for the spacelike geodesics as denoted by red color in
\cref{spacelike-normal} is given by \eq{spacelike-geod}. The total
length of these geodesics can be obtained as
\begin{equation}\label{spacelike-length}
  L^{\rm spacelike}_A=2 T\ell\int_{\epsilon}^{\infty}\frac{du}{2u}
  \frac{1}{\sqrt{u^2+\frac{T^2}{4}}}
  =2\ell\log{\frac{T}{\epsilon}}.
		\end{equation}
		
Now for anomalous contribution, the tangent vector and parallel
transported normal frame vectors along this spacelike geodesic is
given by (in $(t,x,u)$ convention)
 \be\label{spacelike-parallel}
 v^\mu=-\frac{2u}{T\ell}(u,0,t), \quad q^\mu=-\frac{2u}{T\ell}(t,0,u),
 \quad \tilde{q}^{\mu} =(0,\frac{u}{\ell},0).
 \ee
These vectors satisfy the condition given in \eq{spacelike-condition}.
While the normal vector $n$ can be readily fixed at the boundary along
the time direction as described in \cite{Castro:2014tta}, doing so at
the geodesic's other end in the bulk is not straightforward because of
the non-covariance of bulk TMG, requiring a particular
prescription. The normal vectors at the endpoints of the relevant
geodesics is determined as follows.  First, we choose an orthonormal
triad $(v^\mu,n^\mu,\tilde{n}^\mu)$ at the initial endpoint of
interval where $v$ represents the tangent vector, $n$ is aligned with
the boundary future time direction and $\tilde{n}$ is determined by
orthogonality and the right hand convention. For the other endpoint of
spacelike geodesic I as depicted in \cref{spacelike-normal}, the
vector $\tilde{n}$ is maintained as perpendicular to the plane (along
$x$ direction) and $n$ is subsequently determined based on the tangent
vector $v$ and $\tilde{n}$. Secondly, at the final endpoint of the
interval, an orthonormal triad is established with $n$ remaining along
the future time direction and $\tilde{n}$ derived from $n$ and $v$
which is now oriented towards the negative x direction. This specific
$\tilde{n}$ is then utilized at the other endpoint of spacelike
geodesic II to determine the normal vector $n$ using $v$ and
$\tilde{n}$. Finally, these determined normal vectors at the endpoints
of the spacelike geodesics in the bulk are smoothly connected across
null infinity to find the normal vectors $n$ at the endpoints of the
timelike geodesic.

For the spacelike
geodesic I as shown in \cref{spacelike-normal} which starts from
$t=-T/2$ at the boundary ($u\to 0$) and goes to infinity $u\to
\infty$, it can be 
parameterised by $\beta$ as $t=-T/2\cosh\beta$ and $u=T/2\sinh \beta$
such that $\beta=0$ corresponds to the boundary and $\beta\to \infty$
for the other end point of the geodesic.
The normal vector $n$ is
timelike for the case of spacelike geodesic as required by condition
\eq{spacelike-normal-vector-bdy}. Since the other end point is at
infinity and normal vector becomes null at $\beta \to \infty$, we
regularize it by taking some finite but large value of $\beta$ so that
it remains timelike near infinity and then take $\beta \to \infty$ at
the end of computations.
The boundary values of the (normalised) normal vector $n$ can be written as
\begin{equation}\label{n-I}
  n^i_\mu = \frac{\ell}{u_\epsilon}(- 1,0,0),
  \quad n^f_\mu=\frac{\ell}{u_\infty}(- \cosh \beta, 0, \sinh\beta),
\end{equation}
where the initial value of normal vector at boundary is same as
considered in \cite{Castro:2014tta} and determined by the boundary CFT
data. The final value of $n$ is perpendicular to geodesic at null
infinity (as the geodesic approaches the
line $t=-u$ at $u\to\infty$).

Similarly for curve II, we choose the boundary conditions on $n$ as follows
\begin{equation}\label{n-II}
  n^i_\mu = \frac{\ell}{u_\infty}(-\cosh\beta,0,-\sinh\beta),
  \quad n_\mu^f= \frac{\ell}{u_\epsilon}(-1,0,0),
\end{equation}
where the curve II is parametrized by $\beta$ as $t=T/2\cosh\beta$ and
$u=T/2\sinh \beta$ such that $\beta=0$ corresponds to the boundary and
$\beta \to \infty$ denotes the other end point of geodesic II in the
bulk.  Note that these vectors are normalised with respect to the bulk
metric.

Upon using \eq{spacelike-parallel}, \eq{n-I} and \eq{n-II} in
\eq{spacelike-anomalous-formula}, we find that the total anomalous
contribution from spacelike geodesics I and II vanishes i.e
\begin{equation}\label{spacelike-anomaly-geodesic}
     S_{\rm anom}^{\rm spacelike}=0.
\end{equation}
This is very similar to the case for the entanglement entropy for an
unboosted interval
(i.e a pure spacelike interval) \cite{Castro:2014tta}
where anomalous contribution also vanishes.
Therefore, the contribution from
the spacelike geodesics to the on-shell action \eq{spacelike-EE} is
given by
\begin{equation}\label{EE-spacelike-geodesic}
    S_A^{\rm spacelike}=\frac{\ell}{2G_N^{(3)}}\log\frac{T}{\epsilon}.
\end{equation}

\subsection*{Timelike geodesic contribution}

\begin{figure}[H]
	\centering
        \includegraphics[scale=1.2]{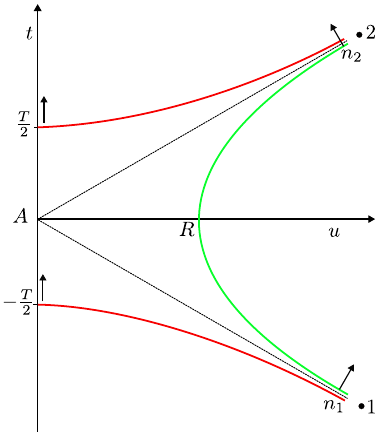}
	\caption{Schematic for the boundary value of normal vectors
          $n_1$ and $n_2$ which are spacelike for timelike
          geodesics.}\label{timelike-normal}
\end{figure}

For the timelike geodesic (green color curve in
\cref{timelike-normal}) as described by \eq{timelike-geod}, the length
of this geodesic can be obtained as
\begin{equation}\label{Length-timelike}
     L^{\rm timelike}_A=2 i\ell R\int_{R}^{\infty}\frac{du}{u}
     \frac{1}{\sqrt{u^{2}-R^{2}}}=i\ell\pi.
		\end{equation}
Now to find the anomalous contribution, the tangent vector and
parallel transported vector to this timelike geodesic satisfying the
condition \eq{timelike-anomalous-formula} can be written as
\be\label{timelike-parallel}
v^\mu=\frac{u}{\ell R}(u,0,t), \quad
q^\mu=-\frac{u}{\ell R}(t,0,u), \quad
\tilde{q}^{\mu} =(0,\frac{u}{\ell},0).
 \ee
The timelike geodesic \eq{timelike-geod} can be parameterized by
$\gamma$ as $u=R\cosh \gamma, t=R \sinh \gamma$ where $\gamma$ range
from $-\infty$ to $+\infty$.  Given that the timelike geodesic
matches up with the spacelike geodesics I, II asymptotically
at the null infinities, we propose to specify the initial
and final normal vectors $n_1$, $n_2$ by matching them up with the
normal vectors $n_i$ and $n_f$ at the respective null infinities (point 1 and 2
in \cref{timelike-normal}).
As such we have
\begin{equation}\label{timelike-n1}
    n^1_{\mu}=\frac{\ell}{u_{-\infty}}(\sinh \gamma,0,\cosh\gamma),
\end{equation}
where $\gamma$ goes to $-\infty$ near the initial point 1. As the
endpoints extend to infinity, we introduce a regularization for
computational purposes by choosing $\gamma$ a large value similar to
the previous subsection. This ensures the boundary value of normal
vectors remain spacelike near infinity i.e at the intersection of
spacelike and timelike geodesics. Similarly we have
\begin{equation}\label{timelike-n2}
     n^2_{\mu}=\frac{\ell}{u_{\infty}}(-\sinh \gamma,0,-\cosh\gamma),
\end{equation}
with $\gamma\to \infty$
for  the normal vector near the point 2 at the null
infinity.
Now using
\eq{timelike-parallel}, \eq{timelike-n1} and \eq{timelike-n2} in
\eq{timelike-anomalous-formula}, we obtain the anomalous contribution
from timelike geodesic as
\begin{equation}\label{S-anom-timelike}
\begin{aligned}
 S_{\rm anom}^{\rm timelike} &= \frac{1}{4G_N^{(3)} \mu}\log
 \left[\frac{q(\tau_2)\cdot n_2-i \tilde{q}(\tau_2)\cdot n_2}{q(\tau_1)\cdot
     n_1-i \tilde{q}(\tau_1)\cdot n_1}\right]\\
& =\frac{1}{4G_N^{(3)} \mu}i\pi.
  \end{aligned}
\end{equation}
Hence the total contribution
to the on-shell action \eq{timelike-onshell-action} for timelike
geodesic can be obtained using \eqref{Length-timelike} and
\eqref{S-anom-timelike} as
\begin{equation}\label{EE-timelike-geodesic}
     S_A^{\rm timelike}=\frac{\ell}{4G_N^{(3)}}i \pi
     +\frac{1}{4G_N^{(3)} \mu}i\pi.
\end{equation}
Now upon using \eq{EE-spacelike-geodesic} and
\eq{EE-timelike-geodesic} in \eq{TEE-total-expression}, we obtain the
holographic TEE in the presence of gravitational anomaly after using
\eq{BH-formula} as follows
\begin{equation}
    S_A^T=\frac{c_L+c_R}{6}\log \frac{T}{\epsilon}+\frac{c_R}{6}i \pi.
\end{equation}
Interestingly, we observe that the above result matches exactly with the dual
field theory result \eqref{TEE-zero-temp}.

\subsection{Holographic TEE at finite temperature and angular potential}

We now compute the holographic timelike entanglement entropy for a
pure timelike interval in a dual CFT$_2$ at finite temperature and
angular potential. The bulk dual for this case is described by the
rotating BTZ black hole (black brane) with the metric
\begin{equation}\label{Rotating-BTZ}
  ds^2 = -\frac{(r^2 - r_+^2)(r^2 - r_-^2)}{r^2 \ell^2} d\tau^2
  + \frac{r^2 \ell^2}{(r^2 - r_+^2)(r^2 - r_-^2)} dr^2
  + r^2 (d\phi + \frac{r_+ r_-}{r^2 \ell}d\tau)^2  .
\end{equation}
where the coordinate $\phi$ is non-compact, and $r_{\pm}$ are the
radii of the outer and inner horizons respectively.  The left and
right moving temperatures corresponding to the left and the right
moving modes in a dual CFT$_{2}$ can be written in terms of horizon
radius as
\begin{equation}\label{Eff-temperature-horizon-rad}
T_{L(R)}=\frac{1}{\beta_{L(R)}}=\frac{r_+\mp r_-}{2\pi \ell}.
\end{equation}
To obtain the geodesic length and anomalous contribution in this
geometry, we can use the fact that the rotating BTZ black hole is
locally equivalent to pure AdS$_3$ so that one can map the metric
\eq{Rotating-BTZ} to the Poincar\'{e} patch of the AdS$_3$ spacetime
as follows
\begin{equation}\label{metric-map}
  x \pm t = \sqrt{\frac{r^2 - r_+^2}{r^2 - r_-^2}} e^{2\pi (\phi \pm \tau)/\beta_{R,L}},
  \,\, \qquad u = \sqrt{\frac{r_+^2 - r_-^2}{r^2 - r_-^2}}
  e^{(\phi r_+ + \tau r_-)/\ell}.
\end{equation}
Then the metric of AdS$_3$ in the Poincar\'{e} coordinates is given by
\begin{equation}\label{Poinc-patch}
 ds^2=\frac{\ell^2}{u^2} (-dt^2+dx^2+du^2).
\end{equation}

Now consider a pure timelike interval $A$ of length $T$ having end
points $(\tau_1,\phi_1)=(-\frac{T}{2},0)$ and
$(\tau_2,\phi_2)=(\frac{T}{2},0)$ in a dual CFT$_2$ at finite
temperature and angular potential. The corresponding end points of the
interval $A$ in the Poincar\'e coordinates i.e $(t_1,x_1)$ and
$(t_2,x_2)$ can be obtained using the map \eq{metric-map}. The
equations for the geodesic to compute the holographic TEE
corresponding to this timelike interval can be written as
\begin{equation}
\begin{aligned}
    \lambda^2 - u^2 &= \frac{R_P^2}{4},\\
    u^2-\lambda^2&=R_0^2
    \end{aligned}
\end{equation}
where $\lambda$ parametrizes the geodesics, mirroring the similar role
$t$ played in \eq{geodesic-zero-T}. The first equation represents the
spacelike geodesics joining end points of interval and null infinities
while the second equation is the timelike geodesic which joins these two
null infinities. Here $R_0$ is some constant and $R_P$ is the length
of timelike interval $A$ in the Poincar\'e coordinates as
\begin{equation}
R_P= \sqrt{(t_2 - t_1)^2-(x_2 - x_1)^2}.
\end{equation}
Now, the total length of these geodesics in Poincar\'e coordinates can be
obtained as
\begin{equation}\label{L-T-BTZ}
    L=\log \frac{R_P^2}{\epsilon_1 \epsilon_2}+i\pi,
\end{equation}
where the relation between the cut-off $\epsilon_{i}$ ($i=1,2$) and
$r_\infty$ for the bulk in the Poincar\'{e} coordinates and rotating
BTZ respectively, can be written using \eq{metric-map} as
\begin{equation}\label{cut-off-relation}
\varepsilon_{i}= \sqrt{\frac{r_{+}^2-r_{-}^2}{r_{\infty}^2}}e^{\tau_i
  r_{-}/\ell}.
\end{equation}
Upon utilizing the coordinate transformation \eq{metric-map} and
\eq{cut-off-relation} in \eq{L-T-BTZ}, we obtain the length
contribution to holographic TEE in terms of the original rotating BTZ
coordinates as
\begin{equation}\label{length-contribution-T}
  S_{\rm EH}=\frac{1}{4G_N^{(3)}} \log \left
  ( \frac{\beta_L \beta_R}{\pi^2 \epsilon^2} \sinh \frac{\pi T}{\beta_L}
  \sinh \frac{\pi T}{\beta_R}\right ) +\frac{1}{4G_N^{(3)}} i \pi,
\end{equation}
where $\epsilon$ is the UV cut-off for the dual CFT$_2$ and related to the
bulk infra red cut-off $r_\infty$ as $\epsilon={\ell}/{r_\infty}$.

For the anomalous contribution, we can compute it separately for
contribution from spacelike and timelike geodesics, however doing it
is very complicated in rotating BTZ background. To compute it more
simply, we use the fact that geodesic comprising of spacelike and
timelike geodesic forms a single continuous curve homologous to the
boundary interval and the integral $\int_C \tilde{n} \cdot \nabla n$
is insensitive to variation of $n$ along the curve and, only depends
on the initial and final value of normal vector $n$. Since the initial
and final points of geodesics lie on the spacelike geodesic, the
boundary conditions on the normal vector $n$ can be written in terms of
time coordinate used to describe the black hole similar to
\cite{Castro:2014tta}
\begin{equation}\label{BC-n-BH}
n_i = n_f = \left(\partial_{\tau}\right)_{\rm CFT}.
\end{equation}
The parallel transported vectors along the spacelike geodesics in the
Poincar\'e coordinates can be obtained by boosting
\eq{spacelike-parallel} as\footnote{Note that the Lorentz boosts
preserve the causal structure of the metric ensuring that timelike
intervals (geodesics) remain timelike and spacelike intervals
(geodesics) remain spacelike.}
\begin{equation}\label{q-btz}
q^{\mu}=-\frac{2u}{\ell R_P}\left(\frac{\lambda
  t_{21}}{R_P},\frac{\lambda x_{21}}{R_P},u\right), \,\,\quad
\tilde{q}^{\mu}=\frac{u}{\ell}\left(\frac{x_{21}}{R_P},\frac{t_{21}}{R_P},0\right).
\end{equation}
Here $\lambda$ plays the role of $t$ as discussed earlier and takes
value at initial point $\lambda_i = -\frac{R_P}{2}$ and $\lambda_f=
\frac{R_P}{2}$ at the final point.  Now using the coordinate
transformation \eq{metric-map} to transform $(q,\tilde{q})$ into the
BTZ coordinates followed by \eq{BC-n-BH} and
\eq{spacelike-anomalous-formula}, we obtain the anomaly contribution
to holographic TEE as follows

\begin{equation}\label{anomaly-contribution-T}
S_{\rm anom} = \frac{1}{4G_N^{(3)} \mu} \log \left ( \frac{\sinh
  \frac{\pi T}{\beta_R}\beta_R }{\sinh \frac{\pi T}{\beta_L}
  \beta_L}\right )+\frac{1}{4G_N^{(3)} \mu}i \pi.
\end{equation}
We observe that it is complex and the imaginary part is similar to the
zero temperature case.  The holographic TEE at finite temperature with
a conserved angular momentum can now be written using
\eq{length-contribution-T}, \eq{anomaly-contribution-T} and
\eq{BH-formula} as follows
\begin{equation}
S_A^T =\frac{c_L+c_R}{12} \log \left ( \frac{\beta_L \beta_R}{\pi^2
  \epsilon^2} \sinh \frac{\pi T}{\beta_L} \sinh \frac{\pi
  T}{\beta_R}\right ) - \frac{(c_L-c_R)}{12} \log \left (\frac{\beta_R \sinh
  \frac{\pi T}{\beta_R} }{  \beta_L\sinh \frac{\pi T}{\beta_L}
}\right )+\frac{c_R}{6}i \pi.
\end{equation}
We again see that the above result matches exactly with corresponding
dual field theory result described in \eq{TEE-finite-T-A}.

\subsection{Holographic TEE at zero temperature and finite angular potential}

We now turn our attention to compute the holographic TEE in CFT$_{2}$
with a conserved charge which is dual to the extremal rotating BTZ
black hole. The metric for the extremal BTZ black hole can be obtained
by taking the extremal limit $(r_{+} = r_{-}=r_0)$ in
\eq{Rotating-BTZ} as follows
\begin{equation}\label{extremal-BH}
 ds^{2}=-\frac{(r^2-r_{0}^2)^2}{r^2 \ell^2}d\tau^2 + \frac{r^2 \ell^2
 }{(r^2-r_{0}^2)^2}dr^2+ r^2(d\phi+\frac{r_{0}^2}{r^2 \ell}d\tau)^2.
\end{equation}
In the extremal limit, the Hawking temperature $T_H$ and left moving
temperature $T_L$ vanishes but there remains an effective temperature
called the Frolov-Thorne temperature that represents the degeneracy of
the ground state of an extremal BTZ black hole and given by
\cite{PhysRevD.39.2125}
\begin{equation}
   T_{FT}=\frac{r_0}{\pi \ell}=\frac{1}{\beta_R}.
\end{equation}
The metric in \eq{extremal-BH} can be mapped to the pure AdS$_3$
spacetime in the Poincar\'e coordinates \eq{Poinc-patch} by the
following coordinate transformations \cite{Keski-Vakkuri:1998gmz}
\begin{equation}\label{Coord-extremal}
    \begin{aligned}
        x+t& =\frac{\ell}{2r_0}e^{\frac{2r_0}{\ell}(\phi+\tau)}, \quad
        \quad x-t=\phi-\tau-\frac{\ell
          r_0}{r^2-r_0^2}\\ u&=\frac{\ell}{\sqrt{r^2-r_0^2}}e^{r_0(\phi+\tau)/\ell},
    \end{aligned}
\end{equation}

Consider a pure timelike interval $A$ of length $T$ having end points
$(\tau_1,\phi_1)=(-\frac{T}{2},0)$ and
$(\tau_2,\phi_2)=(\frac{T}{2},0)$ in a dual CFT$_2$ at zero
temperature and finite angular potential. The computation of the
length of the spacelike and timelike geodesic is similar to the
previous subsection involving the non-extremal black hole. So, the
contribution from length of geodesics to the holographic TEE for this
case can be obtained using \eq{Coord-extremal} in \eq{L-T-BTZ} as
follows
\begin{equation}\label{length-contribution-extremal}
  S_{\rm EH}=\frac{1}{4G_N^{(3)}} \log \left ( \frac{\ell T}{r_0
    \epsilon^2} \sinh \frac{r_0 T}{\ell}\right ) +\frac{1}{4G_N^{(3)}}
  i \pi,
\end{equation}
where $\epsilon$ is the UV cut-off. Similarly we can obtain the
anomalous contribution by transforming the parallel transported
vectors \eq{q-btz} using \eq{Coord-extremal} and subsequently using
\eq{BC-n-BH} in \eq{spacelike-anomalous-formula} which is given by

\begin{equation}\label{anomaly-contribution-extremal}
S_{\rm anom} = \frac{1}{4G_N^{(3)} \mu} \log \left ( \frac{\ell \sinh
  \frac{r_0 T}{\ell}}{r_0 T}\right )+\frac{1}{4G_N^{(3)} \mu}i \pi.
\end{equation}
The holographic TEE at zero temperature with a conserved angular
momentum may now be obtained upon using
\eq{length-contribution-extremal}, \eq{anomaly-contribution-extremal}
and \eq{BH-formula} as follows
\begin{equation}
\begin{aligned}
S_A^T &=\frac{c_L+c_R}{12} \log \left ( \frac{T \, \ell}{r_0
  \epsilon^2} \sinh \frac{r_0 T}{\ell}\right ) - \frac{(c_L-c_R)}{12}
\log \left ( \frac{\ell \sinh \frac{r_0 T}{\ell}}{r_0 T}\right
)+\frac{c_R}{6}i \pi\\ &=\frac{c_L}{6}\log
\frac{T}{\epsilon}+\frac{c_R}{6}\log \left( \frac{\ell}{r_0
  \epsilon}\sinh \frac{r_0 T}{\ell}\right)+\frac{c_R}{6}i \pi.
\end{aligned}
\end{equation}
We observe that the imaginary part corresponding to left moving mode
is absent similar to earlier results and terms containing $c_R$
resembles the TEE at an effective Frolov-Thorne temperature
\cite{PhysRevD.39.2125}. It also matches with the dual field theory
result given in \eq{TEE-zero-T-finite-ap}.

\section{Summary and discussion}\label{sec-summary}

To summarize, we have studied the effect of gravitational anomaly on
the timelike entanglement entropy (TEE) in the context of
AdS$_3$/CFT$_2$ duality. Specifically we employed analytic continuation
and computed the TEE for a pure
timelike interval in a CFT$_2$ with gravitational anomaly in the
vacuum state of CFT$_2$, for thermal CFT$_2$'s with an angular
potential and also for CFT$_2$'s at finite angular potential with zero
temperature.
As a result, the TEE receives
an additional
purely imaginary contribution due to the gravitational
anomaly. Interestingly for a pure timelike interval in all cases, the
imaginary part depends on only one central charge in contrast to its
real counterpart which
depends on both the left and right central charges.
This chiral
sensitivity of the TEE  suggests that it may be used 
as a probe for gravitational anomalies.
We see that in the absence of the gravitational anomaly, our results reduce to the known results in the CFT$_2$.

Subsequently we computed the timelike entanglement entropy for a pure
timelike interval in CFT$_2$ with gravitational anomalies dual to bulk
topologically massive gravity (TMG) in asymptotically AdS$_3$
geometries.  The bulk three dimensional action for the TMG-AdS$_3$
geometries involve a gravitational Chern-Simons term in addition to
the usual Einstein-Hilbert term. In the locally AdS$_3$ geometry we
considered, the worldlines of massive spinning particles are
geodesics. Since the holographic TEE consists of spacelike and
timelike geodesics, it involves the on-shell actions of massive
spinning particles moving on these extremal worldlines (geodesics) in
the dual bulk geometry. We derived from the Chern-Simons action the
anomalous contribution specifically arising from timelike geodesic
which represents the rotation of normal frames along it. Therefore,
the holographic TEE involves the sum of length of the geodesics and
the twist (rotation) of the normal frame along these geodesics. We
further proposed a prescription for the boundary condition on normal
vectors to compute the Chern-Simons contribution to the holographic
TEE. Utilizing this prescription, we determined the holographic TEE in
a CFT$_2$ dual to pure AdS$_3$ and also for the cases of rotating
extremal and non-extremal BTZ black holes. Interestingly we observed
that the holographic results in all these cases agrees precisely with
the dual field theory results.

There are several interesting directions to follow from our work. In
this paper, we obtained the holographic TEE using spacelike and
timelike geodesics.
Exploring the same using the RT surface in a
complexified geometry as proposed in \cite{Heller:2024whi} can offer
new insights about the anomaly contribution. Furthermore, exploring
TEE in the Chern-Simons formulation of $(2+1)$-dimensional
topologically massive gravity theories \cite{Ammon:2013hba} and
geometries that deviate from locally AdS$_3$ such as warped AdS
\cite{Detournay:2012pc} would be an interesting areas of
investigation. Determining the various inequalities satisfied by TEE similar to the entanglement entropy remains an important open problem.
Computing TEE in higher dimensional theories with
gravitational anomalies will also be interesting. Finally it would be
also interesting to investigate TEE in the context of boundary
Lifshitz field theory \cite{Chu:2024nwf} along the lines of AdS/BCFT
\cite{Chu:2023zah}.  These investigations can provide further new
insights into black hole interiors and bulk reconstruction. We leave
these interesting issue for future investigations.

\section*{Acknowledgments}

We thank Dimitrios Giataganas for helpful discussions. H.P would like
to thank Gaurav Katoch, Debajyoti Sarkar and Bhim Sen for
collaboration during the initial stages of this work.  C.S.C
acknowledge support of this work by NCTS and the grant
113-2112-M-007-039-MY3 of the National Science and Technology Council
of Taiwan. H.P acknowledges the support of this work by NCTS.

\bibliographystyle{JHEP}

\bibliography{TEE-a}  

\providecommand{\href}[2]{#2}\begingroup\raggedright\begin{thebibliography}{10}

\bibitem{Calabrese:2004eu}
P.~Calabrese and J.L.~Cardy, \emph{{Entanglement entropy and quantum field
  theory}}, \href{https://doi.org/10.1088/1742-5468/2004/06/P06002}{\emph{J.
  Stat. Mech.} {\bfseries 0406} (2004) P06002}
  [\href{https://arxiv.org/abs/hep-th/0405152}{{\ttfamily hep-th/0405152}}].

\bibitem{Calabrese:2009qy}
P.~Calabrese and J.~Cardy, \emph{{Entanglement entropy and conformal field
  theory}}, \href{https://doi.org/10.1088/1751-8113/42/50/504005}{\emph{J.
  Phys. A} {\bfseries 42} (2009) 504005}
  [\href{https://arxiv.org/abs/0905.4013}{{\ttfamily 0905.4013}}].

\bibitem{Alvarez-Gaume:1983ihn}
L.~Alvarez-Gaume and E.~Witten, \emph{{Gravitational Anomalies}},
  \href{https://doi.org/10.1016/0550-3213(84)90066-X}{\emph{Nucl. Phys. B}
  {\bfseries 234} (1984) 269}.

\bibitem{Castro:2014tta}
A.~Castro, S.~Detournay, N.~Iqbal and E.~Perlmutter, \emph{{Holographic
  entanglement entropy and gravitational anomalies}},
  \href{https://doi.org/10.1007/JHEP07(2014)114}{\emph{JHEP} {\bfseries 07}
  (2014) 114} [\href{https://arxiv.org/abs/1405.2792}{{\ttfamily 1405.2792}}].

\bibitem{He:2023cju}
P.-Z.~He and H.-Q.~Zhang, \emph{{Revisit the entanglement entropy with
  gravitational anomaly}},
  \href{https://doi.org/10.1007/JHEP11(2023)142}{\emph{JHEP} {\bfseries 11}
  (2023) 142} [\href{https://arxiv.org/abs/2305.05892}{{\ttfamily
  2305.05892}}].

\bibitem{Basu:2022nds}
D.~Basu, H.~Parihar, V.~Raj and G.~Sengupta, \emph{{Entanglement negativity,
  reflected entropy, and anomalous gravitation}},
  \href{https://doi.org/10.1103/PhysRevD.105.086013}{\emph{Phys. Rev. D}
  {\bfseries 105} (2022) 086013}
  [\href{https://arxiv.org/abs/2202.00683}{{\ttfamily 2202.00683}}].

\bibitem{Wen:2022jxr}
Q.~Wen and H.~Zhong, \emph{{Covariant entanglement wedge cross-section,
  balanced partial entanglement and gravitational anomalies}},
  \href{https://doi.org/10.21468/SciPostPhys.13.3.056}{\emph{SciPost Phys.}
  {\bfseries 13} (2022) 056}
  [\href{https://arxiv.org/abs/2205.10858}{{\ttfamily 2205.10858}}].

\bibitem{Ryu:2006bv}
S.~Ryu and T.~Takayanagi, \emph{{Holographic derivation of entanglement entropy
  from AdS/CFT}},
  \href{https://doi.org/10.1103/PhysRevLett.96.181602}{\emph{Phys. Rev. Lett.}
  {\bfseries 96} (2006) 181602}
  [\href{https://arxiv.org/abs/hep-th/0603001}{{\ttfamily hep-th/0603001}}].

\bibitem{Ryu:2006ef}
S.~Ryu and T.~Takayanagi, \emph{{Aspects of Holographic Entanglement Entropy}},
  \href{https://doi.org/10.1088/1126-6708/2006/08/045}{\emph{JHEP} {\bfseries
  08} (2006) 045} [\href{https://arxiv.org/abs/hep-th/0605073}{{\ttfamily
  hep-th/0605073}}].

\bibitem{Jiang:2019qvd}
H.~Jiang, \emph{{Anomalous Gravitation and its Positivity from Entanglement}},
  \href{https://doi.org/10.1007/JHEP10(2019)283}{\emph{JHEP} {\bfseries 10}
  (2019) 283} [\href{https://arxiv.org/abs/1906.04142}{{\ttfamily
  1906.04142}}].

\bibitem{Wen:2024muv}
Q.~Wen, M.~Xu and H.~Zhong, \emph{{Timelike and gravitational anomalous
  entanglement from the inner horizon}},
  \href{https://arxiv.org/abs/2412.21058}{{\ttfamily 2412.21058}}.

\bibitem{Kraus:2005zm}
P.~Kraus and F.~Larsen, \emph{{Holographic gravitational anomalies}},
  \href{https://doi.org/10.1088/1126-6708/2006/01/022}{\emph{JHEP} {\bfseries
  01} (2006) 022} [\href{https://arxiv.org/abs/hep-th/0508218}{{\ttfamily
  hep-th/0508218}}].

\bibitem{Song:2016gtd}
W.~Song, Q.~Wen and J.~Xu, \emph{{Modifications to Holographic Entanglement
  Entropy in Warped CFT}},
  \href{https://doi.org/10.1007/JHEP02(2017)067}{\emph{JHEP} {\bfseries 02}
  (2017) 067} [\href{https://arxiv.org/abs/1610.00727}{{\ttfamily
  1610.00727}}].

\bibitem{Jiang:2017ecm}
H.~Jiang, W.~Song and Q.~Wen, \emph{{Entanglement Entropy in Flat Holography}},
  \href{https://doi.org/10.1007/JHEP07(2017)142}{\emph{JHEP} {\bfseries 07}
  (2017) 142} [\href{https://arxiv.org/abs/1706.07552}{{\ttfamily
  1706.07552}}].

\bibitem{Hijano:2017eii}
E.~Hijano and C.~Rabideau, \emph{{Holographic entanglement and Poincar\'e
  blocks in three-dimensional flat space}},
  \href{https://doi.org/10.1007/JHEP05(2018)068}{\emph{JHEP} {\bfseries 05}
  (2018) 068} [\href{https://arxiv.org/abs/1712.07131}{{\ttfamily
  1712.07131}}].

\bibitem{Wen:2018mev}
Q.~Wen, \emph{{Towards the generalized gravitational entropy for spacetimes
  with non-Lorentz invariant duals}},
  \href{https://doi.org/10.1007/JHEP01(2019)220}{\emph{JHEP} {\bfseries 01}
  (2019) 220} [\href{https://arxiv.org/abs/1810.11756}{{\ttfamily
  1810.11756}}].

\bibitem{Gao:2019vcc}
B.~Gao and J.~Xu, \emph{{Holographic entanglement entropy in AdS3/WCFT}},
  \href{https://doi.org/10.1016/j.physletb.2021.136647}{\emph{Phys. Lett. B}
  {\bfseries 822} (2021) 136647}
  [\href{https://arxiv.org/abs/1912.00562}{{\ttfamily 1912.00562}}].

\bibitem{Song:2019txa}
W.~Song and J.~Xu, \emph{{Structure Constants from Modularity in Warped CFT}},
  \href{https://doi.org/10.1007/JHEP10(2019)211}{\emph{JHEP} {\bfseries 10}
  (2019) 211} [\href{https://arxiv.org/abs/1903.01346}{{\ttfamily
  1903.01346}}].

\bibitem{Sun:2008uf}
J.-R.~Sun, \emph{{Note on Chern-Simons Term Correction to Holographic
  Entanglement Entropy}},
  \href{https://doi.org/10.1088/1126-6708/2009/05/061}{\emph{JHEP} {\bfseries
  05} (2009) 061} [\href{https://arxiv.org/abs/0810.0967}{{\ttfamily
  0810.0967}}].

\bibitem{Basu:2021axf}
D.~Basu, A.~Chandra, H.~Parihar and G.~Sengupta, \emph{{Entanglement Negativity
  in Flat Holography}},
  \href{https://doi.org/10.21468/SciPostPhys.12.2.074}{\emph{SciPost Phys.}
  {\bfseries 12} (2022) 074}
  [\href{https://arxiv.org/abs/2102.05685}{{\ttfamily 2102.05685}}].

\bibitem{Wang:2023jfr}
X.-S.~Wang and J.-q.~Wu, \emph{{An observable in Classical Pure AdS$_{3}$
  Gravity: the twist along a geodesic}},
  \href{https://doi.org/10.1007/JHEP05(2024)111}{\emph{JHEP} {\bfseries 05}
  (2024) 111} [\href{https://arxiv.org/abs/2312.10751}{{\ttfamily
  2312.10751}}].

\bibitem{Doi:2022iyj}
K.~Doi, J.~Harper, A.~Mollabashi, T.~Takayanagi and Y.~Taki,
  \emph{{Pseudoentropy in dS/CFT and Timelike Entanglement Entropy}},
  \href{https://doi.org/10.1103/PhysRevLett.130.031601}{\emph{Phys. Rev. Lett.}
  {\bfseries 130} (2023) 031601}
  [\href{https://arxiv.org/abs/2210.09457}{{\ttfamily 2210.09457}}].

\bibitem{Doi:2023zaf}
K.~Doi, J.~Harper, A.~Mollabashi, T.~Takayanagi and Y.~Taki, \emph{{Timelike
  entanglement entropy}},
  \href{https://doi.org/10.1007/JHEP05(2023)052}{\emph{JHEP} {\bfseries 05}
  (2023) 052} [\href{https://arxiv.org/abs/2302.11695}{{\ttfamily
  2302.11695}}].

\bibitem{Nakata:2020luh}
Y.~Nakata, T.~Takayanagi, Y.~Taki, K.~Tamaoka and Z.~Wei, \emph{{New
  holographic generalization of entanglement entropy}},
  \href{https://doi.org/10.1103/PhysRevD.103.026005}{\emph{Phys. Rev. D}
  {\bfseries 103} (2021) 026005}
  [\href{https://arxiv.org/abs/2005.13801}{{\ttfamily 2005.13801}}].

\bibitem{Mollabashi:2020yie}
A.~Mollabashi, N.~Shiba, T.~Takayanagi, K.~Tamaoka and Z.~Wei, \emph{{Pseudo
  Entropy in Free Quantum Field Theories}},
  \href{https://doi.org/10.1103/PhysRevLett.126.081601}{\emph{Phys. Rev. Lett.}
  {\bfseries 126} (2021) 081601}
  [\href{https://arxiv.org/abs/2011.09648}{{\ttfamily 2011.09648}}].

\bibitem{Mollabashi:2021xsd}
A.~Mollabashi, N.~Shiba, T.~Takayanagi, K.~Tamaoka and Z.~Wei, \emph{{Aspects
  of pseudoentropy in field theories}},
  \href{https://doi.org/10.1103/PhysRevResearch.3.033254}{\emph{Phys. Rev.
  Res.} {\bfseries 3} (2021) 033254}
  [\href{https://arxiv.org/abs/2106.03118}{{\ttfamily 2106.03118}}].

\bibitem{Li:2022tsv}
Z.~Li, Z.-Q.~Xiao and R.-Q.~Yang, \emph{{On holographic time-like entanglement
  entropy}}, \href{https://doi.org/10.1007/JHEP04(2023)004}{\emph{JHEP}
  {\bfseries 04} (2023) 004}
  [\href{https://arxiv.org/abs/2211.14883}{{\ttfamily 2211.14883}}].

\bibitem{Basak:2023otu}
J.K.~Basak, A.~Chakraborty, C.-S.~Chu, D.~Giataganas and H.~Parihar,
  \emph{{Massless Lifshitz field theory for arbitrary z}},
  \href{https://doi.org/10.1007/JHEP05(2024)284}{\emph{JHEP} {\bfseries 05}
  (2024) 284} [\href{https://arxiv.org/abs/2312.16284}{{\ttfamily
  2312.16284}}].

\bibitem{Afrasiar:2024lsi}
M.~Afrasiar, J.K.~Basak and D.~Giataganas, \emph{{Timelike entanglement entropy
  and phase transitions in non-conformal theories}},
  \href{https://doi.org/10.1007/JHEP07(2024)243}{\emph{JHEP} {\bfseries 07}
  (2024) 243} [\href{https://arxiv.org/abs/2404.01393}{{\ttfamily
  2404.01393}}].

\bibitem{Afrasiar:2024ldn}
M.~Afrasiar, J.K.~Basak and D.~Giataganas, \emph{{Holographic Timelike
  Entanglement Entropy in Non-relativistic Theories}},
  \href{https://arxiv.org/abs/2411.18514}{{\ttfamily 2411.18514}}.

\bibitem{Heller:2024whi}
M.P.~Heller, F.~Ori and A.~Serantes, \emph{{Geometric Interpretation of
  Timelike Entanglement Entropy}},
  \href{https://doi.org/10.1103/PhysRevLett.134.131601}{\emph{Phys. Rev. Lett.}
  {\bfseries 134} (2025) 131601}
  [\href{https://arxiv.org/abs/2408.15752}{{\ttfamily 2408.15752}}].

\bibitem{Wang:2018jva}
P.~Wang, H.~Wu and H.~Yang, \emph{{Fix the dual geometries of $T\bar{T}$
  deformed CFT$_2$ and highly excited states of CFT$_2$}},
  \href{https://doi.org/10.1140/epjc/s10052-020-08680-7}{\emph{Eur. Phys. J. C}
  {\bfseries 80} (2020) 1117}
  [\href{https://arxiv.org/abs/1811.07758}{{\ttfamily 1811.07758}}].

\bibitem{Liu:2022ugc}
B.~Liu, H.~Chen and B.~Lian, \emph{{Entanglement entropy of free fermions in
  timelike slices}},
  \href{https://doi.org/10.1103/PhysRevB.110.144306}{\emph{Phys. Rev. B}
  {\bfseries 110} (2024) 144306}
  [\href{https://arxiv.org/abs/2210.03134}{{\ttfamily 2210.03134}}].

\bibitem{Guo:2022jzs}
W.-z.~Guo, S.~He and Y.-X.~Zhang, \emph{{Constructible reality condition of
  pseudo entropy via pseudo-Hermiticity}},
  \href{https://doi.org/10.1007/JHEP05(2023)021}{\emph{JHEP} {\bfseries 05}
  (2023) 021} [\href{https://arxiv.org/abs/2209.07308}{{\ttfamily
  2209.07308}}].

\bibitem{Narayan:2022afv}
K.~Narayan, \emph{{de Sitter space, extremal surfaces, and time entanglement}},
  \href{https://doi.org/10.1103/PhysRevD.107.126004}{\emph{Phys. Rev. D}
  {\bfseries 107} (2023) 126004}
  [\href{https://arxiv.org/abs/2210.12963}{{\ttfamily 2210.12963}}].

\bibitem{He:2023eap}
S.~He, J.~Yang, Y.-X.~Zhang and Z.-X.~Zhao, \emph{{Pseudoentropy for descendant
  operators in two-dimensional conformal field theories}},
  \href{https://doi.org/10.1103/PhysRevD.109.025014}{\emph{Phys. Rev. D}
  {\bfseries 109} (2024) 025014}
  [\href{https://arxiv.org/abs/2301.04891}{{\ttfamily 2301.04891}}].

\bibitem{Jiang:2023ffu}
X.~Jiang, P.~Wang, H.~Wu and H.~Yang, \emph{{Timelike entanglement entropy and
  TT\textasciimacron{} deformation}},
  \href{https://doi.org/10.1103/PhysRevD.108.046004}{\emph{Phys. Rev. D}
  {\bfseries 108} (2023) 046004}
  [\href{https://arxiv.org/abs/2302.13872}{{\ttfamily 2302.13872}}].

\bibitem{Narayan:2023ebn}
K.~Narayan and H.K.~Saini, \emph{{Notes on time entanglement and
  pseudo-entropy}},
  \href{https://doi.org/10.1140/epjc/s10052-024-12855-x}{\emph{Eur. Phys. J. C}
  {\bfseries 84} (2024) 499}
  [\href{https://arxiv.org/abs/2303.01307}{{\ttfamily 2303.01307}}].

\bibitem{Chu:2023zah}
C.-S.~Chu and H.~Parihar, \emph{{Time-like entanglement entropy in AdS/BCFT}},
  \href{https://doi.org/10.1007/JHEP06(2023)173}{\emph{JHEP} {\bfseries 06}
  (2023) 173} [\href{https://arxiv.org/abs/2304.10907}{{\ttfamily
  2304.10907}}].

\bibitem{Jiang:2023loq}
X.~Jiang, P.~Wang, H.~Wu and H.~Yang, \emph{{Timelike entanglement entropy in
  dS$_{3}$/CFT$_{2}$}},
  \href{https://doi.org/10.1007/JHEP08(2023)216}{\emph{JHEP} {\bfseries 08}
  (2023) 216} [\href{https://arxiv.org/abs/2304.10376}{{\ttfamily
  2304.10376}}].

\bibitem{He:2023wko}
S.~He, J.~Yang, Y.-X.~Zhang and Z.-X.~Zhao, \emph{{Pseudo entropy of primary
  operators in $ T\overline{T}/J\overline{T} $-deformed CFTs}},
  \href{https://doi.org/10.1007/JHEP09(2023)025}{\emph{JHEP} {\bfseries 09}
  (2023) 025} [\href{https://arxiv.org/abs/2305.10984}{{\ttfamily
  2305.10984}}].

\bibitem{Chen:2023eic}
D.~Chen, X.~Jiang and H.~Yang, \emph{{Holographic TT\textasciimacron{} deformed
  entanglement entropy in dS3/CFT2}},
  \href{https://doi.org/10.1103/PhysRevD.109.026011}{\emph{Phys. Rev. D}
  {\bfseries 109} (2024) 026011}
  [\href{https://arxiv.org/abs/2307.04673}{{\ttfamily 2307.04673}}].

\bibitem{He:2023ubi}
P.-Z.~He and H.-Q.~Zhang, \emph{{Holographic timelike entanglement entropy from
  Rindler method*}},
  \href{https://doi.org/10.1088/1674-1137/ad57a8}{\emph{Chin. Phys. C}
  {\bfseries 48} (2024) 115113}
  [\href{https://arxiv.org/abs/2307.09803}{{\ttfamily 2307.09803}}].

\bibitem{Narayan:2023zen}
K.~Narayan, \emph{{Further remarks on de Sitter space, extremal surfaces, and
  time entanglement}},
  \href{https://doi.org/10.1103/PhysRevD.109.086009}{\emph{Phys. Rev. D}
  {\bfseries 109} (2024) 086009}
  [\href{https://arxiv.org/abs/2310.00320}{{\ttfamily 2310.00320}}].

\bibitem{Shinmyo:2023eci}
K.~Shinmyo, T.~Takayanagi and K.~Tasuki, \emph{{Pseudo entropy under joining
  local quenches}}, \href{https://doi.org/10.1007/JHEP02(2024)111}{\emph{JHEP}
  {\bfseries 02} (2024) 111}
  [\href{https://arxiv.org/abs/2310.12542}{{\ttfamily 2310.12542}}].

\bibitem{Guo:2023tjv}
W.-z.~Guo, Y.-z.~Jiang and Y.~Jiang, \emph{{Pseudo entropy and
  pseudo-Hermiticity in quantum field theories}},
  \href{https://doi.org/10.1007/JHEP05(2024)071}{\emph{JHEP} {\bfseries 05}
  (2024) 071} [\href{https://arxiv.org/abs/2311.01045}{{\ttfamily
  2311.01045}}].

\bibitem{Diaz:2023npx}
N.L.~Diaz, J.M.~Matera and R.~Rossignoli, \emph{{Spacetime quantum and
  classical mechanics with dynamical foliation}},
  \href{https://doi.org/10.1103/PhysRevD.109.105008}{\emph{Phys. Rev. D}
  {\bfseries 109} (2024) 105008}
  [\href{https://arxiv.org/abs/2311.06486}{{\ttfamily 2311.06486}}].

\bibitem{Kanda:2023jyi}
H.~Kanda, T.~Kawamoto, Y.-k.~Suzuki, T.~Takayanagi, K.~Tasuki and Z.~Wei,
  \emph{{Entanglement phase transition in holographic pseudo entropy}},
  \href{https://doi.org/10.1007/JHEP03(2024)060}{\emph{JHEP} {\bfseries 03}
  (2024) 060} [\href{https://arxiv.org/abs/2311.13201}{{\ttfamily
  2311.13201}}].

\bibitem{He:2023syy}
S.~He, Y.-X.~Zhang, L.~Zhao and Z.-X.~Zhao, \emph{{Entanglement and pseudo
  entanglement dynamics versus fusion in CFT}},
  \href{https://doi.org/10.1007/JHEP06(2024)177}{\emph{JHEP} {\bfseries 06}
  (2024) 177} [\href{https://arxiv.org/abs/2312.02679}{{\ttfamily
  2312.02679}}].

\bibitem{Das:2023yyl}
A.~Das, S.~Sachdeva and D.~Sarkar, \emph{{Bulk reconstruction using timelike
  entanglement in (A)dS}},
  \href{https://doi.org/10.1103/PhysRevD.109.066007}{\emph{Phys. Rev. D}
  {\bfseries 109} (2024) 066007}
  [\href{https://arxiv.org/abs/2312.16056}{{\ttfamily 2312.16056}}].

\bibitem{Grieninger:2023knz}
S.~Grieninger, K.~Ikeda and D.E.~Kharzeev, \emph{{Temporal entanglement entropy
  as a probe of renormalization group flow}},
  \href{https://doi.org/10.1007/JHEP05(2024)030}{\emph{JHEP} {\bfseries 05}
  (2024) 030} [\href{https://arxiv.org/abs/2312.08534}{{\ttfamily
  2312.08534}}].

\bibitem{Guo:2024lrr}
W.-z.~Guo, S.~He and Y.-X.~Zhang, \emph{{Relation between timelike and
  spacelike entanglement entropy}},
  \href{https://arxiv.org/abs/2402.00268}{{\ttfamily 2402.00268}}.

\bibitem{Basu:2024bal}
D.~Basu and V.~Raj, \emph{{Reflected entropy and timelike entanglement in
  TT\textasciimacron{}-deformed CFT2s}},
  \href{https://doi.org/10.1103/PhysRevD.110.046009}{\emph{Phys. Rev. D}
  {\bfseries 110} (2024) 046009}
  [\href{https://arxiv.org/abs/2402.07253}{{\ttfamily 2402.07253}}].

\bibitem{Guo:2024edr}
W.-z.~Guo, Y.-z.~Jiang and J.~Xu, \emph{{Pseudoentropy sum rule by analytical
  continuation of the superposition parameter}},
  \href{https://doi.org/10.1007/JHEP11(2024)069}{\emph{JHEP} {\bfseries 11}
  (2024) 069} [\href{https://arxiv.org/abs/2405.09745}{{\ttfamily
  2405.09745}}].

\bibitem{Anegawa:2024kdj}
T.~Anegawa and K.~Tamaoka, \emph{{Black hole singularity and timelike
  entanglement}}, \href{https://doi.org/10.1007/JHEP10(2024)182}{\emph{JHEP}
  {\bfseries 10} (2024) 182}
  [\href{https://arxiv.org/abs/2406.10968}{{\ttfamily 2406.10968}}].

\bibitem{Goswami:2024vfl}
K.~Goswami, K.~Narayan and G.~Yadav, \emph{{No-boundary extremal surfaces in
  slow-roll inflation and other cosmologies}},
  \href{https://doi.org/10.1007/JHEP03(2025)193}{\emph{JHEP} {\bfseries 03}
  (2025) 193} [\href{https://arxiv.org/abs/2409.14208}{{\ttfamily
  2409.14208}}].

\bibitem{Jena:2024tly}
S.S.~Jena and S.~Mahapatra, \emph{{A note on the holographic time-like
  entanglement entropy in Lifshitz theory}},
  \href{https://doi.org/10.1007/JHEP01(2025)055}{\emph{JHEP} {\bfseries 01}
  (2025) 055} [\href{https://arxiv.org/abs/2410.00384}{{\ttfamily
  2410.00384}}].

\bibitem{Xu:2024yvf}
J.~Xu and W.-z.~Guo, \emph{{Imaginary part of timelike entanglement entropy}},
  \href{https://doi.org/10.1007/JHEP02(2025)094}{\emph{JHEP} {\bfseries 02}
  (2025) 094} [\href{https://arxiv.org/abs/2410.22684}{{\ttfamily
  2410.22684}}].

\bibitem{Roychowdhury:2025ukl}
D.~Roychowdhury, \emph{{Holographic timelike entanglement and $c$ theorem for
  supersymmetric QFTs in ($ 0+1 $)d}},
  \href{https://arxiv.org/abs/2502.10797}{{\ttfamily 2502.10797}}.

\bibitem{Guo:2025pru}
W.-z.~Guo and J.~Xu, \emph{{A duality of Ryu-Takayanagi surfaces inside and
  outside the horizon}},  \href{https://arxiv.org/abs/2502.16774}{{\ttfamily
  2502.16774}}.

\bibitem{Roychowdhury:2025aye}
D.~Roychowdhury, \emph{{Field theory aspects of $ \eta $-deformed superstring
  background}},  \href{https://arxiv.org/abs/2503.06294}{{\ttfamily
  2503.06294}}.

\bibitem{Milekhin:2025ycm}
A.~Milekhin, Z.~Adamska and J.~Preskill, \emph{{Observable and computable
  entanglement in time}},  \href{https://arxiv.org/abs/2502.12240}{{\ttfamily
  2502.12240}}.

\bibitem{Alvarez-Gaume:1984zlq}
L.~Alvarez-Gaume and P.H.~Ginsparg, \emph{{The Structure of Gauge and
  Gravitational Anomalies}},
  \href{https://doi.org/10.1016/0003-4916(85)90087-9}{\emph{Annals Phys.}
  {\bfseries 161} (1985) 423}.

\bibitem{Bardeen:1984pm}
W.A.~Bardeen and B.~Zumino, \emph{{Consistent and Covariant Anomalies in Gauge
  and Gravitational Theories}},
  \href{https://doi.org/10.1016/0550-3213(84)90322-5}{\emph{Nucl. Phys. B}
  {\bfseries 244} (1984) 421}.

\bibitem{Hubeny:2007xt}
V.E.~Hubeny, M.~Rangamani and T.~Takayanagi, \emph{{A Covariant holographic
  entanglement entropy proposal}},
  \href{https://doi.org/10.1088/1126-6708/2007/07/062}{\emph{JHEP} {\bfseries
  07} (2007) 062} [\href{https://arxiv.org/abs/0705.0016}{{\ttfamily
  0705.0016}}].

\bibitem{Caputa:2013lfa}
P.~Caputa, V.~Jejjala and H.~Soltanpanahi, \emph{{Entanglement entropy of
  extremal BTZ black holes}},
  \href{https://doi.org/10.1103/PhysRevD.89.046006}{\emph{Phys. Rev. D}
  {\bfseries 89} (2014) 046006}
  [\href{https://arxiv.org/abs/1309.7852}{{\ttfamily 1309.7852}}].

\bibitem{PhysRevD.39.2125}
V.P.~Frolov and K.S.~Thorne, \emph{Renormalized stress-energy tensor near the
  horizon of a slowly evolving, rotating black hole},
  \href{https://doi.org/10.1103/PhysRevD.39.2125}{\emph{Phys. Rev. D}
  {\bfseries 39} (1989) 2125}.

\bibitem{Tachikawa:2006sz}
Y.~Tachikawa, \emph{{Black hole entropy in the presence of Chern-Simons
  terms}}, \href{https://doi.org/10.1088/0264-9381/24/3/014}{\emph{Class.
  Quant. Grav.} {\bfseries 24} (2007) 737}
  [\href{https://arxiv.org/abs/hep-th/0611141}{{\ttfamily hep-th/0611141}}].

\bibitem{Skenderis:2009nt}
K.~Skenderis, M.~Taylor and B.C.~van Rees, \emph{{Topologically Massive Gravity
  and the AdS/CFT Correspondence}},
  \href{https://doi.org/10.1088/1126-6708/2009/09/045}{\emph{JHEP} {\bfseries
  09} (2009) 045} [\href{https://arxiv.org/abs/0906.4926}{{\ttfamily
  0906.4926}}].

\bibitem{Brown:1986nw}
J.D.~Brown and M.~Henneaux, \emph{{Central Charges in the Canonical Realization
  of Asymptotic Symmetries: An Example from Three-Dimensional Gravity}},
  \href{https://doi.org/10.1007/BF01211590}{\emph{Commun. Math. Phys.}
  {\bfseries 104} (1986) 207}.

\bibitem{Hotta:2008yq}
K.~Hotta, Y.~Hyakutake, T.~Kubota and H.~Tanida, \emph{{Brown-Henneaux's
  Canonical Approach to Topologically Massive Gravity}},
  \href{https://doi.org/10.1088/1126-6708/2008/07/066}{\emph{JHEP} {\bfseries
  07} (2008) 066} [\href{https://arxiv.org/abs/0805.2005}{{\ttfamily
  0805.2005}}].

\bibitem{Compere:2008cv}
G.~Compere and S.~Detournay, \emph{{Semi-classical central charge in
  topologically massive gravity}},
  \href{https://doi.org/10.1088/0264-9381/26/1/012001}{\emph{Class. Quant.
  Grav.} {\bfseries 26} (2009) 012001}
  [\href{https://arxiv.org/abs/0808.1911}{{\ttfamily 0808.1911}}].

\bibitem{Lewkowycz:2013nqa}
A.~Lewkowycz and J.~Maldacena, \emph{{Generalized gravitational entropy}},
  \href{https://doi.org/10.1007/JHEP08(2013)090}{\emph{JHEP} {\bfseries 08}
  (2013) 090} [\href{https://arxiv.org/abs/1304.4926}{{\ttfamily 1304.4926}}].

\bibitem{Keski-Vakkuri:1998gmz}
E.~Keski-Vakkuri, \emph{{Bulk and boundary dynamics in BTZ black holes}},
  \href{https://doi.org/10.1103/PhysRevD.59.104001}{\emph{Phys. Rev. D}
  {\bfseries 59} (1999) 104001}
  [\href{https://arxiv.org/abs/hep-th/9808037}{{\ttfamily hep-th/9808037}}].

\bibitem{Ammon:2013hba}
M.~Ammon, A.~Castro and N.~Iqbal, \emph{{Wilson Lines and Entanglement Entropy
  in Higher Spin Gravity}},
  \href{https://doi.org/10.1007/JHEP10(2013)110}{\emph{JHEP} {\bfseries 10}
  (2013) 110} [\href{https://arxiv.org/abs/1306.4338}{{\ttfamily 1306.4338}}].

\bibitem{Detournay:2012pc}
S.~Detournay, T.~Hartman and D.M.~Hofman, \emph{{Warped Conformal Field
  Theory}}, \href{https://doi.org/10.1103/PhysRevD.86.124018}{\emph{Phys. Rev.
  D} {\bfseries 86} (2012) 124018}
  [\href{https://arxiv.org/abs/1210.0539}{{\ttfamily 1210.0539}}].

\bibitem{Chu:2024nwf}
C.-S.~Chu, I.G.~Gonzalez and H.~Parihar, \emph{{Holography for boundary
  Lifshitz field theory}},
  \href{https://doi.org/10.1007/JHEP11(2024)158}{\emph{JHEP} {\bfseries 11}
  (2024) 158} [\href{https://arxiv.org/abs/2409.06667}{{\ttfamily
  2409.06667}}].

\end{thebibliography}\endgroup
  
\end{document}